\documentclass[11pt]{article}

\usepackage{graphicx}
 
\usepackage{setspace}

\usepackage{graphicx}
\usepackage{color}
\usepackage{rotating}
\usepackage{amsmath}
\usepackage{amssymb}
\usepackage{float}
\usepackage{url}

\usepackage{array,arydshln}

\usepackage{url} 

\usepackage[update,prepend]{epstopdf}

\newcommand\bc {\mathbf c}
\newcommand\bd {\mathbf d}

\newcommand\bu {\mathbf u}

\newcommand\bx {\mathbf x}

\newcommand\bz {\mathbf z}

\newcommand\bA {\mathbf A}

\newcommand\bC {\mathbf C}

\newcommand\bW {\mathbf W}

\newcommand\indica {\mathbb{I}}

\newcommand\wpe {\widehat{p}}

\newcommand\wy {\widehat{y}}

\newcommand\wF {\widehat{F}}

\newcommand\wQ {\widehat{Q}}

\newcommand\itA {{\mathcal{A}}}

\newcommand\itC {{\mathcal{C}}}

\newcommand\itK {{\mathcal{K}}}

\newcommand\itM {{\mathcal{M}}}

\newcommand\itP {{\mathcal{P}}}

\newcommand\itS {{\mathcal{S}}}

\newcommand\itU {{\mathcal{U}}}

\newcommand\bbe {\mbox{\boldmath $\beta$}}

\newcommand\bbech {\mbox{\scriptsize${\bbe}$}}
\newcommand\bchi {\mbox{\boldmath $\chi$}}

\newcommand\etab {\mbox{\boldmath $\eta$}}

\newcommand\bgama {\mbox{\boldmath $\gamma$}}
\newcommand\bgamach {\mbox{\scriptsize\boldmath $\gamma$}}

\newcommand\bSi {\mbox{\boldmath $\Sigma$}}

\newcommand\wbbe {\widehat{\bbe}}

\newcommand\weps {\widehat{\epsilon}}

\newcommand\wgamma {\widehat{\gamma}}

\newcommand\wbgama {\widehat{\bgama}}
\newcommand\wbgamach {\mbox{\scriptsize$\wbgama$}}
\newcommand\wlam {\widehat{\lambda}}

\newcommand\wmu {\widehat{\mu}}

\newcommand\wnu {\widehat{\nu}}

\newcommand\wvarsigma {\widehat{\varsigma}}
\newcommand\wtheta {\widehat{\theta}}

\newcommand\wpi {\widehat{\pi}}

\newcommand\wDelta {\widehat{\Delta}}

\newcommand\wtbbe {\widetilde{\bbe}}

\newcommand\wtbeta {\widetilde{\beta}}

\newcommand\wtnu {\widetilde{\nu}}

\def\real{\mathbb{R}}

\newcommand{\esp}{\mathbb{E}}
\newcommand{\prob}{\mathbb{P}}

\newcommand{\var}{\mbox{\sc Var}}

\newcommand{\convpp}{ \buildrel{a.s.}\over\longrightarrow}

\newcommand{\convprob  }{ \buildrel{p}\over\longrightarrow}

\newcommand{\convdist}{ \buildrel{D}\over\longrightarrow}

\newcommand{\trasp}{^{\mbox{\footnotesize \sc t}}}

\newcommand{\traspbis}{{\mbox{\footnotesize \sc t}}}

\newcommand\bcero {{\bf{0}}}

\def\dst{\displaystyle}
\def\median {\mathop{\mbox{median}}}

\def\argmin{\mathop{\mbox{argmin}}}

\newcommand\noi{\noindent}
\def\dst{\displaystyle}

\def\square{\ifmmode\sqr\else{$\sqr$}\fi}
\def\sqr{\vcenter{
         \hrule height.1mm
         \hbox{\vrule width.1mm height2.2mm\kern2.18mm
\vrule width.1mm}
         \hrule height.1mm}}

\newcommand{\ese}{\mbox{\scriptsize \sc s}}

\newcommand\doubrob {\mbox{\scriptsize \sc dr}}
\newcommand\conv {\mbox{\scriptsize \sc conv}}

\newcommand\aipw {\mbox{\scriptsize \sc aipw}}
\newcommand\ipw {\mbox{\scriptsize \sc ipw}}

\begin{document}


\title{Robust location estimators  in regression models with  covariates and responses  missing at random}
\author{{ Ana M. Bianco}$^{(a)}$,  
{ Graciela Boente}$^{(a)}$, 
      {Wenceslao Gonz\'alez--Manteiga}$^{(b)}$
{and}\\
{Ana P\'erez--Gonz\'alez}$^{(c)}$ \\
$^{(a)}$  Universidad de Buenos Aires and  CONICET, Argentina \\
$^{(b)}$  Universidad de Santiago de Compostela, Spain \\
$^{(c)}$  Universidad de Vigo, Spain
}
\date{}  
\maketitle

\begin{abstract}
This paper deals with robust marginal estimation under a general regression model when missing data occur in the response and also in some of covariates. The target is a marginal location parameter which is given through an $M-$functional. To obtain robust Fisher--consistent estimators, properly defined marginal distribution function estimators are considered. These estimators avoid the bias  due to missing values  by assuming a missing at random condition.
Three methods are considered to estimate the marginal distribution function which allows to obtain the $M-$location of interest: the well--known inverse probability weighting, a convolution--based method that makes use of the regression model and an augmented inverse probability weighting procedure that prevents against misspecification. The robust proposed estimators and the classical ones are compared through a numerical study under different missing   models including clean and contaminated samples. We illustrate the estimators behaviour under a nonlinear model. A real data set is also analysed.
 \end{abstract}

 \small

\normalsize

\normalsize
\newpage


\section{Introduction}{\label{intro}}

As is well known, the basis for any regression analysis is to record a response variable $y\in\real$ and a covariate vector $\bx \in \real^{d}$ which are linked through the expression 
\begin{equation}
y=\mu(\bx)+\epsilon \;. \label{regresion}
\end{equation}
where generally it is assumed that the error $\epsilon$ is independent of $\bx$. It is worth noticing that in the classical approach, it is usually assumed  that the errors are centred, i.e.,  $\esp(\epsilon_i)=0$ with finite variance $\var\left(\epsilon_i\right)=\sigma_0^2$. In contrast, in a robust framework no moment conditions are required and two branches have been developed. The most studied setting considers that $\epsilon$ has a symmetric distribution $F_0(\cdot/\sigma_0)$ with $\sigma_0$ the unknown scale parameter. In contrast, when the regression errors are skewed, a given family of densities has to be assumed, see, for instance,  Cantoni and Ronchetti (2006) for an approach under a linear regression model with log--Gamma errors. In this paper, $\mu(\cdot)$  is a general regression function which may link the responses with the independent variables linearly, nonlinearly, nonparametrically  or either through a semiparametric model. To perform the analysis  the practitioner records independent copies  $(y_i, \bx_i\trasp)\trasp$, $1\leq i \leq n$, of $(y,\bx\trasp)\trasp$, that is,  $(y_i, \bx_i\trasp)\trasp$ satisfy \eqref{regresion}
with  the errors $\epsilon_{i}$ i.i.d. and independent of $\bx_i$.  

Suppose we are interested in estimating a location parameter  of the response   $y$. In the classical setting, the target parameter is the mean and it is well known that, even when all observations are available, the mean is very sensitive to the presence of outliers in the sample. Just one outlying observation can take this estimator beyond any limit. Forty years after Huber's (1964) seminal paper, robust estimators are a popular  choice that protects against outliers. Among others,  the median or $M-$estimators,  which   are given through a continuous location functional  $T$, have been developed to overcome the mean sensitivity towards atypical data. In this robust context,  the target is now the robust $M-$location functional related to the estimation procedure   defined through a score function.

Beyond the robust point of view, the effect of ignoring missing observations from the analysis is well known. In particular, when only responses are missing, the estimation of the response mean based on the observed data has deserved a lot of attention. Several strategies have been developed to alleviate the effect on the bias of missing values. Some of them use that additional variables with predictive ability are recorded and include
inverse probability weighted (\textsc{ipw}) or regression based procedures.	In addition to missing responses, some covariates may be dropped out making the problem  more complex.   In order to adjust for missing values both in the response or design variables, it would be necessary to extend existing practice so as to take benefit of the predictive capability of the always observed covariates. The marginal estimation task is even more challenging when atypical responses arise in the sample, since the standard procedures based on maximum likelihood  are very sensitive to the occurrence of a few atypical observations.   
 Furthermore, the challenge is even greater if the aim is to obtain a robust estimator which at the same time protects against misspecification in the missing pattern or in the regression model.

As a motivation, we consider the environmental data analysed in Cleveland (1985) related to air quality measurements.  This data set  consists  in 153 observations that include daily record readings of ozone, wind speed and solar radiation. Several authors have described the nonlinear relation between  these variables. For that reason, they have fitted a nonlinear model to  the ozone measurements, over a subset of the data, corresponding to the always observed cases since missing variables occur. The considered  covariates include the wind speed and the solar radiation.  A robust fit for nonlinear models with missing responses was given in Bianco and Spano (2017) who consider an exponential growth model to explain the ozone daily behaviour in terms of wind speed and identify several atypical data. However, the inclusion of a linear component based on the  solar radiation may add valuable information to the analysis of the ozone variation. An appealing characteristic of these data is that not only some of the ozone records are missing, but also the variable solar radiation has  dropouts, while the variable wind speed is completely observed. Hence, the detection of possible atypical observations when  solar radiation is included in the analysis remains open, as well as the estimation of a reliable location marginal parameter.

For that reason, in this paper, we address robust location estimation in the framework of the regression model \eqref{regresion} when  the response and a fixed subset of independent variables are subject to missingness, while  the remaining covariates are always observed.  This situation may arise, for instance, in environmental observational studies as in our example,  in biological essays when some independent variables can be controlled, while others not or in  epidemiological studies where multivariate survival analyses are performed.  The key point is that when treated inappropriately,  missing values among covariates may affect the postulation of an appropriate  model. As mentioned above, the simple method of deleting from the analysis those cases with missing values, either in the response or the independent variables, may produce biased estimators that may lead to wrong conclusions. For a recent discussion, see for instance,  Chen \textsl{et al.} (2008) and  Hristache and Patilea (2017).  In order to avoid these bias problems and to provide a unified approach to handle missing data, Chen \textsl{et al.} (2015)  consider a missing at random model  where  responses and   covariates  are jointly missing.  It is worth to emphasize that instead of the quantile regression model studied  in Chen \textsl{et al.} (2015)  that concerns the conditional distribution, throughout  this paper we use the regression model \eqref{regresion} as a tool to estimate  the marginal location measure of interest. With these ideas in mind,   to obtain robust marginal estimators under this missing scheme, we will consider the propensity model defined in   Chen \textsl{et al.} (2015).

More precisely, throughout this paper, we assume that the  observations consist on the triplets  $\left(y_i, \bx_i\trasp, \delta_i\right)$, $1\le i \le n$, such that $(y_i,\bx_i\trasp)\trasp=(\bz_i^{(m)\traspbis},\bz_i)\trasp$   with  $\bz_i^{(m)}$ the vector containing the variables subject to missingness, $\bz_i$ the $k-$dimensional vector containing   the always observed variables, $1\le k\le d$. Since our goal is to estimate a marginal parameter,   we focus only on the situation in which   $y_i$ is one of the components of $\bz_i^{(m)}$. The presence indicator variable $\delta_i$  is such that $\delta_i=1$ if all the values in  $\bz_i^{(m)}$ are observed and $0$ otherwise.  The missing at random model (\textsc{mar}) assumes that
\begin{equation}
\prob\left(\delta_i=1\vert (y_i,\bx_i) \right)=\prob\left(\delta_i=1\vert \bz_i\right)=p(\bz_i)\,,
 \label{chen15}
 \end{equation}
  allowing to identify the parameter of interest just in terms of the distribution of the   data available at hand. Note that   \eqref{chen15} enables  to deal with   situations  in which  missing values occur only among the responses and also with those cases where data are missing from the response and a subset of covariates. We refer to Chen \textsl{et al.} (2015) for a thorough analysis regarding the missing scenarios modelled with this framework.

The goal of this paper is to introduce, in the context of the regression model \eqref{regresion}, resistant estimators  for the marginal location of $y$, say $\theta=T(F_y)$, where $F_y$ is the distribution of $y$,  when there are missing values both in the responses and in some (but not all)  covariates. When missing data arise only in the responses and all the covariates are fully observed,  median  estimators have been studied in Zhang \textsl{et al.} (2012) and Diaz (2017), while robust $M-$type location procedures   have been considered in Bianco \textsl{et al.} (2010) and Sued and Yohai (2013).   To deal with the situation in which missing covariates may also arise,  we introduce different methods   under \eqref{chen15}.  
The first method to be considered is based on the  \textsc{ipw} approach introduced in Horvitz and Thompson (1952), where each observation is weighted according to the inverse of the estimated probability of dropouts. The second approach extends the ideas given in M\"uller (2009) to a robust setting when also covariates may be missing. To this end, it is necessary to have a robust and strongly consistent   estimator  of the regression function $\mu$ given in  \eqref{regresion} that will allow to estimate the errors distribution and the distribution of $\mu(\bx)$ as well. An estimator of the distribution function of the responses,  $F_y$, is then constructed by considering their convolution. Finally, the obtained  marginal distribution estimator ensures that it is possible to obtain robust estimators of the marginal quantity $T(F_y)$, given through a continuous functional $T$. In particular, in this paper we focus on $M-$location estimates. 

As when estimating the mean, the estimators of the marginal distribution based on inverse propensity score weighting   require a correct postulated propensity model, while those based on the convolution method  require also a correct postulated regression model. For that reason,  another important novelty of the paper is that  we  introduce an augmented inverse probability weighting (\textsc{aipw}) that  prevents against misspecification of the regression model or the dropouts probability while protecting against atypical observations.  As far as we know, when missing data occur on the responses and some (but not all) of the covariates, our $M-$estimators proposal gives the first attempt to obtain valid estimates when atypical observations arise,  either if the model on the regression function or on the missing probability are correct. In this sense, the new estimator copes with two major purposes: to be robust and double protected.

 The paper is organized as follows. Section \ref{sec:marginal} describes some marginal measures of interest to be used in the sequel. The estimators when missing data occur in the responses and some of the covariates are described in Section \ref{sec:covariates}, where also their asymptotic behaviour is studied. The double protected robust estimator is introduced in Section \ref{sec:DR}. A numerical study is carried out in Section \ref{sec:monte} to examine the small sample properties of the proposed procedures  under a nonlinear regression model.  The ozone data set  is analysed in Section \ref{sec:ejemplo}, where the advantage of
the proposed \textsc{aipw} procedure over the convolution--based approach is illustrated, while some  concluding remarks  and recommendations are given in Section \ref{sec:conclusion}.  All proofs are relegated to the Appendix.

\section{Notation and preliminaries}{\label{sec:marginal}}
Throughout this paper, we  denote as $Q_y$ the response marginal  probability measure   and as $F_y(s)=Q_y((-\infty,s])$ its related distribution function. Let $\theta=T(F_y)=T(Q_y)$ be any marginal $M-$location functional, where we will use indistinctly the notation $T(F_y)$ or $T(Q_y)$. Some examples of usual interest are the marginal mean or median of $y_1$ which are special cases of  $M-$functionals.

 Mmany  $M-$estimators are defined using a previously computed nuisance parameter estimator. A typical example consists on the traditional marginal location--scale model, where $F_y$ has  scale $\varsigma_0$. The scale plays the role of the nuisance parameter and to obtain  a scale equivariant procedure,   an initial estimator $\wvarsigma_0$ is needed.

From now on, $\rho$ stands for a \textsl{rho}--function as defined in Maronna \textsl{et al.} (2006, Chapter 2), i.e., a function $\rho$  such that 
\begin{itemize}
\item $\rho(u)$ is a nondecreasing function of  $|u|$, 
\item $\rho(0)=0$,
\item $\rho(u)$ is increasing for  $u>0$ when $\rho(u)<\|\rho\|_{\infty}$,
\item if $\rho$ is bounded, it is also assumed that $\|\rho\|_{\infty}=1$.
\end{itemize} 
The related $M-$location functional $T_\rho(Q_y)$ of $y$ may be defined as 
$$T_\rho(Q_y)=\argmin_{a\in \real}\esp\,\rho\left( \frac{y-a}{S(Q_y)}\right)\,,$$ 
where $S(Q_y)$ is the scale functional to be defined below. Usually, $\rho=\rho_c$ with $\rho_c(u)= \rho^{\star}(u/c)$ where $\rho^{\star}$ is a $\rho-$function and $c > 0$  is a   tuning constant chosen to attain a given efficiency. A common choice for a bounded $\rho-$function is the bisquare Tukey's function $\rho^{\star}(u)=\min\left(3 u ^2 - 3 u ^4 +  u ^6, 1\right)$. For example, the choice $c=4.685$ gives a 95\% of efficiency with respect to the mean under normality.

If $\rho$ is continuously differentiable with  derivative   $\psi=\rho^{\prime}$, when considering the differentiating equations,  one has that $\lambda(T_\rho(Q_y),S(Q_y)) =0$ where $\lambda(a,\varsigma)=\esp\psi\left(({y_1-a})/{\varsigma}\right)$.  In particular, when 
\begin{equation}
\lambda(T_\rho(Q_y),\varsigma) =0\quad \mbox{ for any } \quad \varsigma>0\,,
\label{psi0}
\end{equation}
 the $M-$location estimator distribution is independent of the preliminary scale estimator distribution. 
 
 Two well known examples of $M-$functionals are the mean and the median which correspond to $\rho(u)=u^2$ and $\rho(u)=|u|$, respectively. A feature of the estimators related to these functionals is that, in both cases, it is not necessary to have a preliminary estimation of the nuisance parameter. Moreover, these functionals allow to have a deeper insight on the interpretation of an $M-$location parameter either for symmetric or skewed distributions. When $F_y$ is symmetric around $\theta$, both functionals coincide with $\theta$. Furthermore, we also have that $T_\rho(Q_y)=\theta$ for any $\rho-$function and (\ref{psi0}) holds for any   odd function $\psi$. On the contrary, when  $y$ is skewed, the situation is different. In fact, both functionals are well identified, but they do not coincide; an illustrative example may be the $\chi^2-$distribution.
The same assertion holds for any general $M-$location functional.

Since the scale of a distribution measures its dispersion, it is sensible to choose scale estimators that are invariant under translations and equivariant under scale transformations (see    Maronna \textsl{et al.} 2006). Among other robust scale functionals, common choices are the \textsc{mad} (median of the absolute values around the median) and $S-$dispersion functionals, which are related to $M-$ scale estimators  (see Huber and Ronchetti, 2009).
To define the latter, let $\rho_0$ be a $\rho-$function. One possible choice is  $\rho_{0}(t)=\rho_{c_0}(t)=\rho^{\star}(t/c_0)$, as above, where the user--chosen tuning constant  $c_0 > 0$  guarantees Fisher--consistency under the underlying distribution. The $S-$dispersion functional is then defined as 
\begin{equation}
S(Q_y)=\min_{a} S(Q_y,a)= S(Q_y,\theta_{\ese}(Q_y))\quad \mbox{and}\quad \esp\rho_0\left(\frac{y-a}{S(Q_y,a)}\right)=b\,,
\label{S-est}
\end{equation}
where $0<b<1$   and $\theta_{\ese}(Q_y)$ is usually called the $S-$location functional. 
For instance, when $\rho^{\star}(t)= \indica_{|t|>1}$ and $b=1/2$, $S(Q_y,a)= \median(|y-a|)/c_0$, leading to the least median location estimator.
As mentioned in Maronna \textsl{et al.} (2006), when $\rho_0$ is bounded, the breakdown point of the $M-$scale
estimator is $\min(b, 1- b)$, since $\| \rho_0 \|_\infty = 1$.  When, as in our simulation study, $\rho^{\star}( y)$ is the Tukey's bisquare function  and we take
$c_0 = 1.54764$ and $b=1/2$, the estimator is Fisher--consistent
at the normal distribution and has breakdown point 50\%.

An estimator  of $\theta= T_\rho(F_y)$ may be obtained from  a random sample $y_1,\dots, y_n$  plugging--in  an estimator of the marginal distribution function $F_y$. When all the observations are available,   the empirical distribution,  $\wF_{y,n}$, can be computed and thus,   the estimator may be defined as $\wtheta=T_\rho(\wF_{y,n})$. 
Hence,   the $M-$location estimator   is the value $\wtheta$ such that
$$\wtheta=\argmin_{a\in \real} \frac 1n \sum_{i=1}^n \rho\left(\frac{y_i-a}{\wvarsigma}\right)\,,$$
where $\wvarsigma$ stands for a robust consistent estimator of the marginal scale of the response variable, such as  $\wvarsigma=S(\wF_{y,n})$ with $S(\cdot)$ defined in (\ref{S-est}) and $\rho(u)\le \rho_0(u)$ for any $u$. In particular, if $\psi=\rho^{\prime}$, we have that
$$  \frac 1n \sum_{i=1}^n \psi\left(\frac{y_i-\wtheta}{\wvarsigma}\right)=0\,.$$

When missing data arise, the estimators described above cannot be computed in practice or will be biased if only the available observations are used. Section  \ref{sec:covariates} describes some alternatives to solve this problem using the information provided by those covariates that are always observed.

\section{Marginal $M-$location estimators when missing covariates and responses arise}{\label{sec:covariates}}

In this section, we face the problem of estimating an $M-$location parameter $\theta=T(F_y)$ under the  regression model (\ref{regresion}),
when missing data arise both on the responses and on some  covariates and when, at the same time, anomalous responses (vertical outliers) occur. 

We will consider  an incomplete data set $\left(y_i, \bx_i\trasp, \delta_i\right)$, $1\le i \le n$, where $(y_i,\bx_i\trasp)\trasp=(\bz_i^{(m)\traspbis},\bz_i)\trasp$ are defined as in Section \ref{intro}.    The binary variable $\delta_i$ is modelled through \eqref{chen15}. As mentioned above, among the missing variables we will always include the responses, i.e.,  $y_i$ is one of the components of $\bz_i^{(m)}$. 

To adapt to the missing values, two estimators of the marginal probability measure $Q_y$ can be defined (see  Bianco \textsl{et al.}, 2018). The first one is an inverse probability weighting estimator that corrects the bias caused in the estimation by the missing mechanism using  an estimator  of the missingness probability $\wpe (\bz)$. The second one  uses the information given by the assumed regression model. For that purpose,  a convolution type estimator, as the one described in M\"uller  (2009)   for a fully parametric model with missing responses, is defined.
 
Denote as $\wF_{y}$ any of these marginal distribution estimators. Then,  an estimator of the $M-$location parameter may be defined as $\wtheta=T(\wF_{y})$. In Section \ref{ws} and \ref{conv}, we give a precise definition of the $M-$location estimators and we study their asymptotic behaviour.
In particular, to derive consistency results for the inverse probability weighting and  the convolution--based $M-$location estimators,  some of the following assumptions will be needed
\begin{itemize}
\item[\textbf{A1}] $\inf_{\bz\in \itS_\bz} p(\bz)=i_p>0$, where $ \itS_\bz$ is the support of the distribution of $\bz_1$.
\item[\textbf{A2}]  $\sup_{\bz\in  \itS_\bz} |\wpe(\bz)-p(\bz) | \convpp 0$.
\item[\textbf{A3}] $\dst\sup_{\bx\in\itK}|\wmu(\bx)-\mu(\bx)|\convpp 0$, for any compact set $\itK\in \real^d$. 
\end{itemize}

\subsection{The inverse probability weighted $M-$estimator}{\label{ws}}
The \textsc{ipw} $M-$estimator exploits the regressors potential to predict the propensity function $p(\bz)$. More precisely, as it usual when dealing with missing data, using the fully observed data  and inverse probability weighting, $Q_y$ can be estimated by
\begin{equation}
\wQ_{y,\ipw}=\frac{1}{\dst\sum_{\ell=1}^n\frac{\delta_{\ell}}{\wpe(\bz_\ell)}} \sum_{i=1}^n\frac{\delta_{i}}{\wpe(\bz_i)} \Delta_{y_i} =\sum_{i=1}^n \tau_i \Delta_{y_i}\;,
\label{Qws}
\end{equation}
where $\Delta_a$ is the point mass at point $a$ and   $\wpe$ is an estimator of the missing probability $p$.  

Theorem 3.1 in Bianco \textsl{et al.} (2018) ensures that, under \textbf{A1} and \textbf{A2},    $\Pi(\wQ_{y,\ipw} ,Q_y)\convpp 0$, where  $\Pi$ stands for the  Prohorov distance. Therefore, for any    functional $T$ continuous  with respect to the Prohorov distance, we have that $T(\wQ_{y,\ipw})\convpp T(Q_y)$. Recall that continuity with respect to $\Pi$ is a usual requirement when considering robust estimators. In particular, let $\wvarsigma$ be a robust consistent estimator of the marginal scale $\varsigma_0$. A possible choice is $\wvarsigma=S(\wQ_{y,\ipw})$ with $S(\cdot)$ the $S-$dispersion functional related to a $\rho-$function $\rho_0$ as defined in (\ref{S-est}). In this case,   we have that  $\wvarsigma \convpp S(Q_y)=\varsigma_0$.

Regarding the $M-$location estimators, let $\rho$ be a $\rho-$function such that $\psi=\rho^{\prime}$ is   bounded. Two possible families for the  function  $\psi$ may be chosen corresponding to increasing  scores  or    redescending ones. For the first family, the related loss function  $\rho$ is a convex  one, such as the well known Huber's function, while for the second one $\rho$ is a bounded function such as the Tukey's bisquare function. In the latter, it is usually required  $\rho(u)\le  \rho_0(u)$, for any $u$, as mentioned in Section \ref{sec:marginal}.   Denote as  $\wtheta_{\ipw}=T_\rho(\wQ_{y,\ipw})$, then $\wtheta_{\ipw} $ is the solution of $\wlam_{\ipw}(\wpe,\wvarsigma, a)=0$ with
\begin{equation}
\wlam_{\ipw}(q,\varsigma,a)=\sum_{i=1}^n\frac{\delta_i}{q(\bz_i)}\psi \left(\frac{y_i-a}{\varsigma}\right)\;.
\label{UN}
\end{equation} 
 For redesceding $\psi$ functions, in order to identify a proper solution, it is better to define $\wtheta_{\ipw}$ as the value $\wtheta_{\ipw}=\argmin_{a}D_n(\wpe,\wvarsigma,a)$, where
\begin{equation*}
D_n(q,\varsigma, a)=\sum_{i=1}^n\frac{\delta_i}{q(\bz_i)}\rho\left(\frac{y_i-a}{\varsigma}\right)\;.
\label{DN}
\end{equation*}
When   $\psi$ is a differentiable function with bounded derivative $\psi^\prime$, such that $\int |\psi^\prime(u)|du<\infty$, standard  arguments allow to conclude that $T_\rho(\wQ_{y,\ipw})\convpp T_\rho(Q_y)$, since the scale estimators are consistent and $\Pi(\wQ_{y,\ipw} ,Q_y)\convpp 0$.
  
\vskip0.1in
Different $M-$location estimators are obtained according to the procedure chosen to estimate the missing probability. Under certain experimental designs, the propensity $p(\bz)$ may be assumed to be known. This can also be though as the case of   an oracle situation. In contrast, if  $p(\bz)$ is unknown, it may be estimated using a either nonparametric  approach  or a   parametric one based on previous information.  

More precisely, when the propensity is fully known, the marginal $M-$estimator denoted $\wtheta_{\ipw}^{(1)}$ solves $\wlam_{\ipw}(p, \wvarsigma, a)=0$, i.e., we have that
\begin{equation}
\wlam_{\ipw}(p, \wvarsigma, \wtheta_{\ipw}^{(1)})=0\,.
\label{theta1ipw}
\end{equation}

When  a  parametric model   $p(\bz)=p(\bz, \bgama_0)$ is assumed for the missing probability, usually the propensity is estimated by plugging--in a consistent estimator of the unknown parameter  $\bgama_0\in \real^s$, where the dimension $s$ may be different from $k$, the dimension of $\bz$. More precisely, let $\wbgama$ be any consistent estimator of $\bgama$, i.e., such that $\wbgama\convpp\bgama _0$. Hence, the estimator of the missingness probability   defined as $\wpe(\bz)= p(\bz, \wbgama)=\wpe_{\wbgamach}(\bz)$ satisfies \textbf{A2} if $p(\bz, \bgama )$ is equicontinuous in $\bgama$ at $\bgama_0$. This condition holds, for instance, when $p(\bz, \bgama )$ is a continuous function of all its arguments and the support $\itS_\bz$ is a bounded set.  Under a parametric model, we denote as   $\wtheta_{\ipw}^{(2)}$   the solution of $\wlam_{\ipw}(\wpe_{\wbgamach}, \wvarsigma, a)=0$, i.e., 
\begin{equation}
\wlam_{\ipw}(\wpe_{\wbgamach}, \wvarsigma, \wtheta_{\ipw}^{(2)})=0\,.
\label{theta2ipw}
\end{equation}

Finally, when considering a nonparametric smoother, a kernel estimator of the propensity may be defined as $\wpe(\bz)=p_{n,b_n}(\bz)$ where
\begin{equation}
p_{n,b_n}(\bz)={\dst\sum_{i=1}^n K \left(\displaystyle\frac{\bz_i-\bz}{b_n}\right)\delta_i} \left\{\displaystyle\sum_{j=1}^n K  \left(\displaystyle\frac{\bz_j-\bz}{b_n}\right)\right\}^{-1}\;,
\label{kerneldelta}
\end{equation}
with $K :\real^{k}\to \real$ a kernel function and  $b_n$   the smoothing parameter. In this case, if $p(\bz)$ is a uniformly continuous function, $b_n\to 0$, $n b_n^{k}/\log(n)\to +\infty$ and $K(\bz)=\itK(\|\bz\|)$, where $\itK:\real_{\ge 0} \to \real_{\ge 0}$ is a bounded variation function, analogous arguments to those considered in Chapter 2 of Pollard (1984) allow to show that \textbf{A2} holds.   We will denote as $\wtheta_{\ipw}^{(3)}$,  the marginal $M-$estimators obtained when using the nonparametric estimator of the missingness probability, that is, $\wtheta_{\ipw}^{(3)}$ solves
\begin{equation}
\wlam_{\ipw}(p_{n,b_n}, \wvarsigma, \wtheta_{\ipw}^{(3)})=0\,,
\label{theta3ipw}
\end{equation} 
and provides strongly consistent estimators, under \textbf{A1} and the conditions on the kernel and bandwidth mentioned above. 

It is worth mentioning that, under \textbf{A1},  \textbf{A2} holds and regularity conditions on the  function $\rho$, $\wtheta_{\ipw}^{(1)}$, $\wtheta_{\ipw}^{(2)}$ and $\wtheta_{\ipw}^{(3)}$ are strongly consistent to $ T_\rho(Q_y)$.

\vskip0.1in
From now on, let $(y,\bx\trasp, \delta)$ be a random vector with the same distribution as $(y_i, \bx_i\trasp, \delta_i)$, where, as above, $(y,\bx\trasp)\trasp=(\bz ^{(m)\traspbis},\bz)\trasp$ . Henceforth, we will denote by $u=(y-\theta)/\varsigma_0$. 

Theorem {\ref{ws}}.1 summarizes the asymptotic  behaviour of $\wtheta_{\ipw}^{(j)}$, $j=1,2,3$, when the missingness probability is either known or estimated under a parametric model or using a kernel approach. To establish their asymptotic distribution assumptions \textbf{N1} to \textbf{N7} given in the Appendix are needed. 

As mentioned in Section \ref{sec:marginal}, for data with no missing observations, if 
\begin{equation}
\label{numerar}
\esp\psi \left(\frac{y-\theta}\varsigma\right)=0 \qquad \qquad \mbox{ for all $\varsigma>0$}\,,
\end{equation}
  the asymptotic distribution of the  $M-$location estimator does not depend on that of the scale estimator and only its consistency  is required. The   inverse probability weighting $M-$estimators have the same behaviour as shown in Theorem   {\ref{ws}}.1. Among other scale consistent estimators, the practitioner may choose the \textsc{mad} of $\wQ_{y,\ipw}$ or more generally, an $S-$dispersion estimator $ S(\wQ_{y,\ipw})$.  Remark \ref{ws}.1 discusses the situation in which  $F_y$ does not satisfy (\ref{numerar}), which includes skewed distributions.

 For simplicity of notation, we denote as $r(\bz)=\esp\left(\psi\left(u\right)|\bz\right)$ and as 
\begin{equation}
\bd=\esp\left(\frac{\dot{p}(\bz,\bgama_0)}{p(\bz,\bgama_0)}\psi\left(u\right)\right)=\esp\left(\frac{\dot{p}(\bz,\bgama_0)}{p(\bz,\bgama_0)}r(\bz) \right)\,,\label{bb}
\end{equation}
where $\dot{p}(\bz,\bgama)$ stands for the gradient  of $p(\bz, \bgama)$ with respect to $\bgama$.

\vskip0.1in
\noindent \textbf{Theorem {\ref{ws}}.1.} \textsl{Assume that  \textbf{A1}, \textbf{N1} and \textbf{N2}   hold. Let  $\theta$ be such that  (\ref{numerar}) holds. Furthermore, let  $\wvarsigma$ be a scale estimator such that  $\wvarsigma\convprob\varsigma_0$. 
\begin{itemize}
\item[a)] Let $\wtheta_{\ipw}^{(1)}$ be defined in (\ref{theta1ipw}). If  $\wtheta_{\ipw}^{(1)}\convprob \theta$, we have that ${\sqrt{n}}(\wtheta_{\ipw}^{(1)}-\theta)\convdist N(0,\varsigma_0^2\,\upsilon_{\ipw}^{(1)})$, 
where
$$\upsilon_{\ipw}^{(1)}={\esp\left(\frac{\psi^2\left(u\right)}{p(\bz)}\right)}\left(\esp\psi^{\prime}\left(u\right)\right)^{-2}= \gamma_{\ipw}^{(1)}\left(\esp\psi^{\prime}\left(u\right)\right)^{-2}\,.$$  
\item[b)] Assume that $p(\bz)=p(\bz, \bgama_0)$, where  $\bgama_0\in \real^s$. Let $\wpe(\bz)= p(\bz, \wbgama)$, with $\wbgama$   an   estimator of $\bgama_0$ such that $\wbgama\convprob\bgama_0$. Denote as $\wtheta_{\ipw}^{(2)}$   the \textsc{ipw} estimator given by (\ref{theta2ipw}).  If, in addition,  $\wtheta_{\ipw}^{(2)}\convprob \theta$ and  \textbf{N3} and \textbf{N4} hold,   we have that
$\sqrt{n}(\wtheta_{\ipw}^{(2)}-\theta)\convdist N(0, \varsigma_0^2\,\upsilon_{\ipw}^{(2)})$, 
where $\upsilon_{\ipw}^{(2)}={\gamma_{\ipw}^{(2)}}\left(E\psi^{\prime}\left(u\right)\right)^{-2}$ with 
\begin{eqnarray*}
\gamma_{\ipw}^{(2)}&=& \esp\left[\frac{\delta}{p(\bz,\bgama_0)}\psi\left(u\right)-\etab(\bz)\trasp \bd\right]^2= \esp\frac{\psi^2\left(u\right)}{p(\bz)}+\bd\trasp \left\{\bSi \; \bd- 2 \esp\left[  \psi\left(u\right) \etab(\bz) \right]\right\}\\
&=& \esp\frac{\psi^2\left(u\right)}{p(\bz)}+\esp\left(\psi\left(u\right) \frac{\dot{p}(\bz,\bgama_0)}{p(\bz,\bgama_0)}\right)\trasp \left\{\bSi \; \esp\left(\psi\left(u\right) \frac{\dot{p}(\bz,\bgama_0)}{p(\bz,\bgama_0)} \right)- 2 \esp\left[  \psi\left(u\right) \etab(\bz) \right]\right\}
\end{eqnarray*}
and $\etab$ and $\bSi$  given in \textbf{N4}. 
\item[c)]  Let $\wpe(\bz)=p_{n,b_n}(\bz)$ be the kernel estimator  defined in (\ref{kerneldelta}) and  $\wtheta_{\ipw}^{(3)}$ be the \textsc{ipw} estimator defined in  (\ref{theta3ipw}).  If, in addition,   $\wtheta_{\ipw}^{(3)}\convprob \theta$  and   \textbf{N5} to \textbf{N7}  hold,  we have that
$\sqrt{n}(\wtheta_{\ipw}^{(3)}-\theta)\convdist N(0, \varsigma_0^2\,\upsilon_{\ipw}^{(3)})$,
where $\upsilon_{\ipw}^{(3)}=\gamma_{\ipw}^{(3)}\left(\esp\psi^{\prime}\left(u\right)\right)^{-2}$ and 
\begin{eqnarray*}
\gamma_{\ipw}^{(3)}&=& \esp\left(\frac{\delta}{p(\bz)}\psi\left(u\right)-\frac{ \left(\delta-p(\bz)\right)}{p(\bz)}\; r(\bz)\right)^2= \esp\left(\frac{\psi^2\left(u\right)}{p(\bz)}\right)- \esp \left(\frac{ 1-p(\bz) }{p(\bz)}\;r^2(\bz)\right)\;.
\end{eqnarray*}
\end{itemize}
}

\vskip0.1in
 \noindent \textbf{Remark \ref{ws}.1.} Even when the propensity is assumed to be known, the efficiency  with respect to the   \textsc{ipw} mean estimator   depends on the proportion of missing data appearing in the sample, an effect that has been already addressed in the literature when missing values arise   only in the responses. Besides, since   $\gamma_{\ipw}^{(3)}\le \gamma_{\ipw}^{(1)}$, we have that  $\upsilon_{\ipw}^{(3)}\le \upsilon_{\ipw}^{(1)}$ and so, the marginal location estimator $\wtheta_{\ipw}^{(3)}$ computed estimating the missing probability through a kernel estimator is more efficient than that computed with the true propensity, $\wtheta_{\ipw}^{(1)}$. As discussed among others in Wang \textsl{et al.} (1997), the better efficiency of the marginal location estimator $\wtheta_{\ipw}^{(3)}$ is related to the sample adjustment obtained through the propensity  kernel estimator.  The reader can find an heuristic justification  of this behaviour for regression   estimators in  Robins \textsl{et al.} (1994) when only covariates are missing.
 
Note that since we focus on marginal measures, we have considered $M-$location estimators to protect against outliers   in the responses. The situation where atypical data   in the covariates used to model the propensity is not considered here and we refer to Molina \textsl{et al.} (2017) for further discussion. In particular, if the propensity is modelled nonparametrically the lack of atypical observations in the covariate space is a usual assumption to avoid isolated points.  

\vskip0.1in
\noindent\textbf{Remark \ref{ws}.2.} When $u$ has  a skewed distribution, one cannot ensure that (\ref{numerar}) holds, which is a condition used to guarantee that only consistency  is required to the scale estimator  $\wvarsigma$.
To solve this problem, a Bahadur expansion for the scale estimators is useful to derive the results. 
For that purpose, assume that the scale is related to an $S-$estimator with $\rho-$function $\rho_0$, i.e., that the scale equals $\wvarsigma=S(\wQ_{y,\ipw})$ with $S(\cdot)$   defined in (\ref{S-est}) and recall that $T_{\rho}(Q_y)=\argmin_{a} \esp \rho((y-a)/S(Q_y))$. Using analogous arguments to those considered in Sued and Yohai (2013), it is easy to see that the asymptotic variance of the estimators can be obtained as in Theorem {\ref{ws}}.1, if we replace  in the  expressions for $\gamma_{\ipw}^{(j)}$, $j=1,2,3$ given above, the function $\psi$   by
$$\chi(y)=\psi\left(\frac{y-T_{\rho}(Q_y)}{\varsigma_0}\right) - \frac{A_{11}}{A_{10}}\,\left\{\rho_0\left(\frac{y-\theta_{\ese}(Q_y)}{\varsigma_0}\right)-b\right\}$$
  with $\theta_{\ese}(Q_y)$ defined in (\ref{S-est}) and
$$A_{11}= \esp\left[ \left(\frac{y-T_{\rho}(Q_y)}{\varsigma_0}\right)\,\psi^{\prime}\left(\frac{y-T_{\rho}(Q_y)}{\varsigma_0}\right)\right]\qquad A_{10}= \esp\left[ \left(\frac{y-\theta_{\ese}(Q_y)}{\varsigma_0}\right)\,\psi\left(\frac{y-\theta_{\ese}(Q_y)}{\varsigma_0}\right)\right]\,.$$

\subsection{The convolution based marginal $M-$estimator}{\label{conv}}
For the situation in which the missing values are restricted to occur only on the responses,  M\"uller (2009) and Sued and Yohai (2013)  noted that  a different estimator of $F_y$ may be obtained using the regression model and the fact that $F_y$ is the convolution of the errors and     the regression function distributions. 
From now on, we denote as $F_{\epsilon}$ and  $F_\mu$ the distribution function   of the errors $\epsilon$ and  of the true regression function $\mu(\bx)$, respectively. The probability measures $Q_{\epsilon}$ and $Q_\mu$ are defined similarly. Using the convolution property, i.e.,   $F_y=F_{\epsilon}*F_\mu$, a consistent estimator for $F_y$ can be obtained plugging--in consistent estimators $\wF_{\epsilon}$ and $\wF_\mu$ of $F_{\epsilon}$ and $F_\mu$, respectively.
More precisely, the \textsl{fully imputed} estimator introduced in M\"uller (2009) was  adapted to the situation of missing values in the responses and covariates   in Bianco \textsl{et al.} (2018) with the purpose of estimating the marginal quantiles. We recall its definition. 

Let $\wmu(\bx)$ be a consistent estimator of $\mu(\bx)$. This consistent estimation can be accomplished in different ways according to  the model structure assumed on the regression function which may be     parametric, nonparametric or semiparametric.   Bianco \textsl{et al.} (2018) illustrates through a detailed discussion   how  classical consistent estimators of $\mu(\bx)$ may be obtained in  different regression scenarios when missing observations occur in the responses and some of the covariates. However, the estimators defined therein are sensitive to atypical observations since they are mainly based on a least squares approach. It is worth to highlight that in our framework, besides consistency, robustness is also a desirable property for the estimators $\wmu(\bx)$.
Remark {\ref{conv}}.1 discusses some robust consistent alternatives when a parametric model is considered. 

Using the robust regression estimator $\wmu(\bx)$, define
\begin{eqnarray}
 {\wQ}_\mu &= &  \frac 1{\dst\sum_{\ell=1}^n \frac{ \delta_{\ell}}{\wpe(\bz_\ell)}}\sum_{i=1}^n \frac{\delta_{i}}{\wpe(\bz_i)} \Delta_{\wmu(\bx_i)} =\sum_{i=1}^n \tau_i \Delta_{\wmu(\bx_i)}\,,\label{wQmu}
 \end{eqnarray}
where the weights $\tau_i$  are normalized  to guarantee that $\wQ_{\mu}$ is a probability measure.  Note that \eqref{wQmu} involves not only the regression estimator $\wmu(\bx)$ but also, due to the missingness of some covariates, a propensity estimator  $\wpe(\bz)$. Hence,  to avoid biases in the  estimation of $Q_{\mu}$, both the regression and the propensity models must be correctly specified.

When $\delta_{i } =1$, the residuals can be effectively predicted as $\weps_i=y_i-\wmu(\bx_i)$, so that an estimator of $Q_{\epsilon}$ can be computed as
$ 
\wQ_{\epsilon}=   \sum_{i=1}^n \kappa_i \Delta_{\weps_i} 
  $, with $\kappa_i= \delta_{i}/ \sum_{\ell=1}^n \delta_{\ell}$. The convolution--based estimator of  $Q_y$ is then defined as 
$\wQ_{y, \conv} = \wQ_{\epsilon} * \wQ_\mu $. As when missing values arise only on the responses,  $\wQ_{y, \conv}$ is a weighted empirical distribution since it can be written as $\wQ_{y, \conv} =  \sum_{i=1}^n   \sum_{j=1}^n  \kappa_i \tau_j\Delta_{\wy_{ij}}$, 
  where $\wy_{ij} = \wmu(\bx_j) + \weps_i$, for $  i, j \in \{\delta_\ell=1\}  $.

Under mild conditions, $\wQ_\mu$ is a consistent estimator of $Q_\mu$, since condition (\ref{chen15}) holds.  More precisely, if \textbf{A1} to \textbf{A3} hold,    Theorem 3.2 in Bianco \textsl{et al.} (2018) entails that  $\Pi(\wQ_{y, \conv} ,Q_y)\convpp 0$, which leads to the strong consistency of $T_\rho( \wF_{y, \conv})$.

As above, let $\wvarsigma$ be a robust consistent estimator of the marginal scale $\varsigma$, for instance, the \textsc{mad} of  $\wQ_{y, \conv}$ or $\wvarsigma=S(\wQ_{y, \conv})$ with $S(Q_y)$ defined in \eqref{S-est}.  Note that $\wtheta_{\conv}=T_\rho( \wF_{y, \conv})$ is the solution of $\wlam_{\conv}(\wpe,\wvarsigma, a)=0$ where 
\begin{equation*}
\wlam_{\conv}(q,\varsigma,a)=\sum_{i=1}^n\sum_{j=1}^n \frac{\delta_i}{q(\bz_i)}\delta_j \psi \left(\frac{\wy_{ij}-a}{\varsigma}\right)\;.
\label{UNCONV}
\end{equation*}
As in Section \ref{ws}, we denote  respectively as $\wtheta_{\conv}^{(1)}$ and $\wtheta_{\conv}^{(2)}$ the convolution--based estimators obtained assuming that the propensity is known ($\wpe\equiv p$) and that the propensity is estimated using a parametric model, that is, $p(\bz)=p(\bz, \bgama_0)$ and $\wpe(\bz)=p(\bz, \wbgama)$  with $\wbgama$   an   estimator of $\bgama_0$.

 Theorem {\ref{conv}}.1 below provides the asymptotic distribution of $\wtheta_{\conv}^{(j)}$, $j=1,2$, when (\ref{numerar}) holds and $\mu(\bx)$ has a parametric form, i.e., when $\mu(\bx)=m(\bx,\bbe_0)$ as stated in assumption \textbf{N8}. Otherwise, when (\ref{numerar}) does not hold, as in Section   \ref{ws}, a Bahadur expansion for $\wvarsigma$ is needed to obtain an expression for the asymptotic variance of  $\wtheta_{\conv}^{(j)}$.

 Let   $C(y_0,\bx_0,\delta_0)= A(\bx_0,\delta_0)+  B(\epsilon_0,\delta_0)+ ({1}/{\varsigma_0})\, \delta_0 \bchi_1(y_0,\bx_0)\trasp \bc$ where $ \bchi_1$ is defined in \textbf{N8} and
\begin{eqnarray}
\bc&=& \esp\left[ \delta_1 \psi^{\prime}\left(\frac{\epsilon_1+\mu(\bx_2)-\theta}{\varsigma_0}\right)\left\{\dot{m}(\bx_2,\bbe_0)-\dot{m}(\bx_1,\bbe_0)\right\}\right]
\nonumber\\
&=& \esp\left[ \delta_1 \psi^{\prime}\left(u_1+\frac{ \mu(\bx_2)-\mu(\bx_1)}{\varsigma_0}\right)\left\{\dot{m}(\bx_2,\bbe_0)-\dot{m}(\bx_1,\bbe_0)\right\}\right]
\nonumber\\
A(\bx_0,\delta_0)&=& \frac{\delta_0}{p(\bz_0)} \esp \left[\delta_1 \psi\left(\frac{\epsilon_1+\mu(\bx_2)-\theta}{\varsigma_0}\right)|\bx_2=\bx_0\right]\nonumber\\
&=&  \frac{\delta_0}{p(\bz_0)} \esp \left[\delta_1 \psi\left(u_1+\frac{ \mu(\bx_0)-\mu(\bx_1)}{\varsigma_0}\right)\right]
\label{Ax0}\\
B(\epsilon_0,\delta_0)&=&  {\delta_0} \esp \left[  \psi\left(\frac{\epsilon_2+\mu(\bx_1)-\theta}{\varsigma_0}\right)|\epsilon_2=\epsilon_0\right]= {\delta_0} \esp \left[  \psi\left(\frac{\epsilon_0+\mu(\bx_1)-\theta}{\varsigma_0}\right)\right] 
\label{Beps0}\,,
\end{eqnarray}
with $\dot{m}(\bx,\bbe)$  the gradient vector  of the function $m(\bx,\bbe)$  with respect to $\bbe$. Furthermore, when the propensity is estimated using the parametric model $p(\bz)=p(\bz, \bgama_0)$, define $D(y_0,\bx_0,\delta_0)= C(y_0,\bx_0,\delta_0)- \esp(\delta) \etab(\bz_0)\trasp \bd_1$ where  $\etab$  is  given in \textbf{N4} and 
\begin{eqnarray}
\bd_1&=&\esp\left(\frac{\dot{p}(\bz,\bgama_0)}{p(\bz,\bgama_0)} r_{1,2}(\bz)  \right)
\nonumber\\
r_{1,2}(\bz)&=&  \esp \left[ \psi\left( \frac{ \epsilon_1+\mu(\bx_2)-\theta}{\varsigma_0}\right)\Big|{\bz_2=\bz}\right]= \esp \left[ \psi\left(u_1+\frac{ \mu(\bx_2)-\mu(\bx_1)}{\varsigma_0}\right)\Big|{\bz_2=\bz}\right]\,.
\nonumber
 \end{eqnarray}

\vskip0.1in
\noindent \textbf{Theorem {\ref{conv}}.1.} \textsl{Let $\theta$ be such that (\ref{numerar}) holds. Assume that   \textbf{A1}, \textbf{N1}, \textbf{N2} and \textbf{N8}  hold, $t\psi^{\prime}(t)$ is bounded    and  $\wvarsigma\convprob\varsigma_0$. 
\begin{itemize}
\item[a)] Denote as $\wtheta_{\conv}^{(1)}$   the solution of $\wlam_{\conv}(p, \wvarsigma, a)=0$. If  $\wtheta_{\conv}^{(1)}\convprob \theta$, we have that ${\sqrt{n}}(\wtheta_{\conv}^{(1)}-\theta)\convdist N(0,\varsigma_0^2\,\upsilon_{\conv}^{(1)})$, 
where $\upsilon_{\conv}^{(1)}= \gamma_{\conv}^{(1)}\left(\esp\psi^{\prime}\left(u_1\right)\right)^{-2}$ 
and $ \gamma_{\conv}^{(1)}=  \esp C^2(y_1,\bx_1,\delta_1)/\esp \delta_1  $. 
\item[b)] Assume that $p(\bz)=p(\bz, \bgama_0)$, where  $\bgama_0\in \real^s$. Let $\wpe(\bz)= p(\bz, \wbgama)=\wpe_{\wbgamach}(\bz)$, with $\wbgama$   an   estimator of $\bgama$ such that $\wbgama\convprob\bgama_0$ and denote as $\wtheta_{\conv}^{(2)}$   the solution of $\wlam_{\conv}(\wpe_{\wbgamach}, \wvarsigma, a)=0$.  If  $\wtheta{\conv}^{(2)}\convprob \theta$ and  \textbf{N3} and \textbf{N4} hold,   we have that
$\sqrt{n}(\wtheta_{\conv}^{(2)}-\theta)\convdist N(0, \varsigma_0^2\,\upsilon_{\conv}^{(2)})$, 
where $\upsilon_{\conv}^{(2)}=\gamma_{\conv}^{(2)}\left(E\psi^{\prime}\left(u\right)\right)^{-2}$ with 
$  \gamma_{\conv}^{(2)}=  \esp D^2(y_1,\bx_1,\delta_1) /\esp \delta_1$. 
 \end{itemize}
}

\vskip0.1in
\noindent \textbf{Remark {\ref{conv}}.1.} An important step in the computation of the convolution based estimators is the estimation of the regression function, that has its own interest. Furthermore,   the distribution of $\wtheta_{\conv}$ depends on that of the estimator $\wbbe$ of $\bbe_0$.  Among other possible regression models, the   linear or nonlinear regression ones provide a wide class of parametric models. Taking into account that, in our framework, the propensity does not depend on the responses and using the \textsc{mar} assumption \eqref{chen15}, it is easy to see that the simplified regression $MM-$estimators considered in Sued and Yohai (2013) can also be considered in our framework. Moreover, the simplified weighted $MM-$estimators defined in Bianco and Spano (2017) can easily be adapted to this setting when the weights controlling leverage points depend only on the fully observed covariates. Even though the simplified estimators are computed with the observations at hand, i.e., with the complete data set only, standard arguments, similar to those considered in  Sued and Yohai (2013), allow to show that, in this case,   $\wbbe$ is consistent and admits a Bahadur expansion as required in \textbf{N8}.

\section{A double protected and robust location estimator}{\label{sec:DR}}
In this section, we introduce an estimator of $Q_y$ that will allow to provide reliable estimates either if the model on the regression function holds or if the model for the missing probability is correct. The proposed estimator is based in the augmented inverse probability weighting (\textsc{aipw}) method that was introduced, in the framework of casual inference, by Robins \textsl{et al.} (1994), Robbins (1999) and Scharfstein \textsl{et al.} (1999), see also Glynn and Quinn  (2010). The \textsc{aipw} estimator has the attractive property that it is consistent whenever at least one of the models, the propensity   or the regression one, is correctly specified. In this sense, \textsc{aipw} estimators are double protected.

Assume that $\wpi(\bz)$ is an estimator of the missing probability when   a model $\pi(\bz)$ for the missing probability is postulated and denote as $\wpi_i=\wpi(\bz_i)$. When missing data arise only on the responses, Wang and Qin (2010) introduced an augmented inverse probability weighted (\textsc{aipw}) estimator by estimating the conditional distribution of $y_1$ given that $\bx_1=\bx$ using a kernel estimator. Their approach can be extended to the present setting in which missing covariates and responses arise, since the \textsc{mar} assumption entails that $G(y|\bz)=\esp \left(\indica(y_1\le y)|\bz_1=\bz\right)=\esp \left(\indica(y_1\le y)|\bz_1=\bz, \delta_1=1\right)$. Hence, the \textsc{aipw} estimator of the marginal distribution is defined as follows
\begin{equation}
\wF_{y,\aipw}(y)= \frac 1n \sum_{i=1}^n\frac{\delta_{i}}{\wpi(\bz_i)} \indica_{\{y_i\le y\}} + \left(1-\frac{\delta_{i}}{\wpi(\bz_i)}\right) G_n(y|\bz_i)\;,
\label{faiwp}
\end{equation}
where
$$G_n(y|\bz)= \left\{ \dst\sum_{\ell=1}^n   K_1 \left(\displaystyle\frac{\bz_\ell-\bz}{a_n}\right) \delta_\ell \right\}^{-1}\,  \dst\sum_{j=1}^n   K_1 \left(\displaystyle\frac{\bz_j-\bz}{a_n}\right) \delta_j  \indica_{\{y_j\le y\}}\;,
 $$
with $K_1 :\real^{k}\to \real$ a kernel function and  $a_n$   the smoothing parameter. Thus, the \textsc{aipw} estimator of the distribution function $F_y$ is the \textsc{ipw} estimator which is augmented with the information that the always observed covariates $\bz$ provide about the outcomes $y$ through a smooth estimator of the conditional distribution $G(y|\bz)$. In this sense, it is expected that this additional information would allow to obtain more accurate estimators than the \textsc{ipw} ones which do not depend on the regression model.

Denote as
\begin{equation}
\zeta_j= \frac{\delta_j}{\wpi_j} \quad \mbox{and} \quad \varpi_{j} = \delta_j\, \sum_{i=1}^n \left(1-\zeta_i\right) \frac{ K_1 \left(\displaystyle\frac{\bz_j-\bz_i}{a_n}\right)}{\sum_{\ell=1}^n K_1 \left(\displaystyle\frac{\bz_\ell-\bz_i}{a_n}\right)\delta_\ell} 
\label{kapa}
\end{equation}
Then, the \textsc{aipw}  marginal distribution estimator  can be written as a weighted empirical distribution
\begin{eqnarray*}
\wF_{y,\aipw}(y)&=&\frac 1n \sum_{j=1}^n\zeta_j \indica_{\{y_j\le y\}}+ \frac 1n \sum_{i=1}^n \left(1-\zeta_i\right)G_n(y|\bz_i)
 =\frac 1n \sum_{j=1}^n( \zeta_j+\varpi_j) \indica_{\{y_j\le y\}}
 \;,
 \end{eqnarray*}
 where the weights $\zeta_j+\varpi_j$ depend on the missing indicator, the propensity estimator and the observed covariates $\bz_i$. 
 
We will consider the following set of assumptions
\begin{itemize}
\item[\textbf{B1}] $\inf_{\bz\in \itS_\bz} \pi(\bz)=i_{\pi}>0$, where $ \itS_\bz$ is the support of the distribution of $\bz_1$
\item[\textbf{B2}]  $\sup_{\bz\in  \itS_\bz} |\wpi(\bz)-\pi(\bz) | \convpp 0$.
\item[\textbf{B3}] $\sup_{y\in \real} \sup_{\bz \in \itC} |G_n(y|\bz)-G(y|\bz)|\convpp 0$ for any compact set $\itC\subset\itS_\bz$.  
\end{itemize}
It is worth noticing that  the arguments used in the proof of Proposition 3.2.1 in Boente \textsl{et al.} (2009) allow to show that $\sup_{y\in \real} \sup_{\bz \in \itC} |G_n(y|\bz)-G(y|\bz)|\convpp 0$, if $a_n\to 0$ and $n a_n^{k}/\log(n)\to +\infty$ and $K_1(\bz) $ is a bounded Lipschitz function.  Hence, assumption \textbf{B3} is fulfilled in this situation. On the other hand, assumptions \textbf{B1} and \textbf{B2} are similar to \textbf{A1} and \textbf{A2} and involve the postulated propensity  and its estimator.
 
 The next theorem shows that the marginal distribution estimators are consistent. 
\vskip0.1in

\noi \textbf{Theorem \ref{sec:DR}.1.} \textsl{Let $\left(y_i,
\bx_i\trasp,  \delta_i\right)$, $1\le i \le n$ be i.i.d. random vectors over $(\Omega, \itA, \prob)$, $\bx_i\in\real^d$,   such that    (\ref{regresion}) and (\ref{chen15}) hold. Assume that \textbf{B1} to \textbf{B3} hold. 
  Then,  we have that    $\|\wF_{y,\aipw} -F_y\|_{\infty}\convpp 0$ and  $\Pi(\wQ_{y,\aipw},Q_y)\convpp 0$, where $\Pi(P_1,P_2)$ stands for the Prohorov distance between $P_1$ and $P_2$.}

\vskip0.2in
The estimator $\wF_{y,\aipw}$ prevents from misspecifications   in the propensity and does not suffer from misspecification of the regression model. Thus, it allows to define marginal estimators that inherit this property. Effectively, if $\zeta_j$ and $\varpi_j$ are defined as in (\ref{kapa}), an $M-$estimator $\wtheta_{\aipw}=T_\rho( \wF_{y, \aipw})$ can be defined as the solution of $\wlam_{\aipw}(\wpi,\wvarsigma, a)=0$ where   
\begin{equation}
\wlam_{\aipw}(\wpi,\varsigma,a)=\frac 1n\sum_{j=1}^n \left[\zeta_j+ \varpi_j\right]  \psi \left(\frac{y_j-a}{\varsigma}\right)\;,
\label{UNDR}
\end{equation}
and $\wvarsigma$ is a previously computed estimator of the scale $\varsigma_0$, such as $\wvarsigma=S(\wQ_{y,\aipw})$ with $S(\cdot)$ given in \eqref{S-est}. 
Similar arguments to those considered in the proof of Theorem 4.1 in Bianco \textsl{et al.} (2010) allow to derive the following result.

\noi \textbf{Theorem \ref{sec:DR}.2} \textsl{Let $\left(y_i,
\bx_i\trasp,  \delta_i\right)$, $1\le i \le n$ be i.i.d. random vectors over $(\Omega, \itA, \prob)$, $\bx_i\in\real^d$,   such that equations   (\ref{regresion}) and (\ref{chen15}) hold. Assume that  \textbf{B1} to \textbf{B3} hold. Let $\psi:\real \to \real$ be a bounded, differentiable function with bounded derivative $\psi^\prime$, such that $\int |\psi^\prime(u)|du<\infty$.
Furthermore, assume that   $\wvarsigma\convpp \varsigma_0$  and that the function $\lambda(a,\varsigma_0)=\esp\psi\left(({y_1-a})/{\varsigma_0}\right)$ has a unique change of sign, in a neighbourhood of $\theta=T_\rho(F_y)$. Then,  there exists a solution $\wtheta_{\aipw}$ of $\wlam_{\aipw}(\wpi,\wvarsigma,a)=0$,  such that $\wtheta_{\aipw}\convpp \theta$.}

\vskip0.1in
\noindent \textbf{Remark \ref{sec:DR}.1.} Note that assumptions \textbf{A2}, \textbf{B2} and \textbf{A3} require consistent estimators of the propensity and the regression function estimators in order to derive consistency results for the three families of marginal $M-$estimators.
In this paper,  we are also concerned about robustness of the marginal estimators, hence when considering the convolution--based estimator, it seems natural to estimate $\mu(\bx)$ in a robust fashion as described in Remark {\ref{conv}}.1. 

\section{Monte Carlo study}{\label{sec:monte}}

In this section, we present the results of a simulation study carried out   to investigate the finite-sample properties of the location estimators proposed in Section \ref{sec:covariates} and \ref{sec:DR}, under  a nonlinear  regression model. The marginal location estimators $\wtheta=T(\wQ_{y})$ compared in this numerical study are the mean, the median and the $M-$location marginal  related to the bisquare function  $\rho(u) = \rho^{\star}(u/c)$  where $\rho^{\star}( u) = \min\left(3 u ^2 - 3 u ^4 +  u ^6, 1\right)$ and  $c=4.685$. The preliminary scale estimator needed for the $M-$location was taken as  an $S-$estimator with $b=0.5$ and computed also using the Tukey's bisquare function  with tuning constant $c = 1.54764$.
In all cases, we carried out $1000$ replications with samples of size $n=100$ and we considered clean    and contaminated samples containing missing data. 

The goal  of this numerical experiment is two--fold, since we are concerned about robustness  and double protection. On one side, as it is usual in robustness, we  aim  to compare  the behaviour of the classical and robust estimators  under contamination and under clean samples,  but in the scenario where missing data arise in the responses and some of the covariates. On the other hand, by computing   the estimators   $\wtheta_{\ipw}$, $\wtheta_{\conv}$ and $\wtheta_{\aipw}$   defined in Sections \ref{ws}, \ref{conv} and \ref{sec:DR},  we are interested in studying  the performance of the three  proposals considered in this paper not only when the regression model and the missing probability are correctly modelled, but also when one of them is  misspecified. Furthermore, our interest is not only to compare the inverse probability weighting procedure, the convolution--based method and the proposed \textsc{aipw} estimator between them, but also with that of the robust estimator  that would be computed if the complete data set were available. Note that this  last estimator, which corresponds to $p(\bz)\equiv 1$, cannot be computed in practice. However, one of our   aims is to seek which of the proposals would give  mean square errors closer to those obtained if  there were no missing values.

As mentioned in Sections \ref{sec:covariates} and \ref{sec:DR}, the estimators $\wtheta_{\ipw}$, $\wtheta_{\conv}$ and $\wtheta_{\aipw}$ depend on the choice of the propensity estimator $\wpe$. For that reason, we also compare the performance of the location estimators when different missingness estimators are considered. First of all, the location estimators are computed assuming that the propensity is  known, i.e.,  $\wpe=p$. Even though this setting may  seem    unrealistic, it is computed for comparison purposes since it   allows to analyse the influence of estimating the propensity on the location estimator. We also use  a parametric model to fit the propensity, that is, $p(\bz)$ is estimated using the true logistic model    generating the missing observations. This case is labelled as  $\wpe=\wpe_{\log}$ in all Tables and Figures. As mentioned in Remark \ref{ws}.1,   the \textsc{ipw}   marginal location estimator   computed estimating the missing probability with a kernel estimator is more efficient than that computed with the true propensity and this fact should be reflected in our numerical results. For that reason, we also consider a kernel estimator   based on  the Epanechnikov kernel with smoothing parameter   chosen using a cross--validation criterion to estimate the propensity. This case  will be denoted as $\wpe=\wpe_{K}$.  
 
 Finally, the   augmented inverse probability weighted   estimator $\wtheta_{\aipw}$ involves an additional smoothing step to estimate the conditional distribution $G(y|\bz)$. The kernel smoother $G_n(y|\bz)$ was computed using a biweight kernel $K_1(t)=15(1-t^2)^2/16 \,\indica_{(-1,1)}(t)$ with bandwidth $a_n=n^{-1/3}$ as suggested in Wang and Qin (2010).

To evaluate the performance of the estimators under  misspecification, we  considered two possible  situations. In the first one, we estimate  the missing probability as if the model were a missing completely at random (\textsc{mcar}) model, i.e., $p(\bz)=p$, instead of the true logistic one that generates the missing variables $\delta_i$. In all Tables and Figures,   $\wpe_c$ corresponds to  the situation where the estimated missing probability is based on a \textsc{mcar}. In the second misspecification case, the regression model was assumed to be linear instead of the true nonlinear one. 

\subsection{Simulation settings}{\label{sec:setting}}

As mentioned above,  we report here  the marginal estimators performance under a nonlinear regression model. We first generate observations such that
\begin{equation}
y_i= \mu(\bx_i) +\epsilon_i = \beta_2 x_{2,i}+  \beta_3 \exp(\beta _1 \, x_{1,i})+\epsilon_i  \quad 1\le i\le n \, ,
\label{expo}
\end{equation}
where $\bx_i=(x_{1,i}, x_{2,i})$, $\bbe_0=(\beta_1,\beta_2,\beta_3)=(2,0.1,5)$.  The errors $\epsilon_i$ are i.i.d. $  N(0,1)$  and independent of the covariates $\bx_i$  in the non--contaminated case, denoted $C_0$, that is the errors scale $\sigma_0$ equals $1$. The distributions of $x_{1,i}$ and $x_{2,i}$ are  $\itU(0,1)$ and $N(0,1)$, respectively.  The considered contamination, denoted $C_1$,   is such that 10\% of the responses are replaced by $2\left(\beta_2 x_{2,i}+  \beta_3 \exp(\beta _1 \, x_{1,i})\right)$ to obtain observations with large residuals, that is, we generate vertical outliers. Even when this contamination scheme does not generate identically distributed observations as in the gross--errors model,  this  kind of outliers are very harmful (see Fasano, 2009, and  Bianco and Spano, 2017) justifying our choice.

It is worth noticing that, even when the regression errors  are normally distributed, under model (\ref{expo})  the marginal distribution of the responses is not symmetric. For that reason, we have computed  the target functionals, corresponding to clean samples, using $100$ replications of samples of size $10^6$. In this way, the approximated marginal values have a standard error smaller than 0.0015.  The obtained values are reported in  Table \ref{tab:verdaderas} and are considered as target quantities when computing the bias and the mean square error of our estimators.


\begin{table}[ht]
\caption{\label{tab:verdaderas} Target marginal values under  the nonlinear model (\ref{expo}).}  
\vskip0.1in
\begin{center}
\begin{tabular}{c ccc }
\hline
&\multicolumn{3}{c }{$C_0$}\\
\hline
$T(F)$& Mean & Median & $M-$est \\ 
\hline
&16.030 & 13.690 & 15.399\\
\hline
\end{tabular}

\end{center}
\end{table}

We consider  the  following missing setting denoted $\itM({H})$. Given a sample following the model (\ref{expo}), we set $(y_i,x_{2,i})$ as missing if $\delta_i=0$, where $\delta_i$ is a Bernoulli variable with success probability  $p(x_1)=\prob\left(\delta_{i}=1\vert x_1\right)=   1/\left\{{1+ \exp\left[-0.2 x_1-0.2\right]}\right\}$.  
Hence, $\bz_i=x_{1,i}$ and $\bz_i^{(m)}=(y_i, x_{2,i})\trasp$. Under $\itM({H})$  the proportion of missing data is around a  $25\%$.

As mentioned above, to have a benchmark allowing to study the loss of the different marginal estimators when missing values occur, we have computed the estimators with the original sample, that is, taking $\prob(\delta_i=1|\bz_i)\equiv 1$. To identify the obtained results, the label $\itM(1)$ is used in all Tables and Figures. 

Taking into account  the \textsc{mar} assumption, the   robust estimator of the regression parameter $\bbe$ may be computed    using a simplified $MM-$estimator which leads to a consistent procedure (see Remark {\ref{conv}}.1).  We choose as $\rho-$function  the bisquare function with tuning constant such that it will achieve 95\% efficiency under normal errors.

\subsection{Simulation results}{\label{sec:resultado}}
We report the bias, standard deviation and mean square error of the considered marginal estimators. Besides,  to evaluate only the effect of the missingness and the advantage of the given methods we compute   two measures which allow to compare the effect on the estimators of both the contamination and the missingness. For simplicity, let $T(F)$ be the functional to be studied and denote as $\wtheta_{j,0}=T(\wQ_{y,n})$ the estimate obtained in the $j-$replication under $C_0$ when all the data are available, i.e., under $\itM(1)$. Furthermore,  for any missing scheme $\itM$, propensity estimator method $\wpe$ and contamination $C_s$, let $\wtheta_{j, \wpe, s}$ be the estimator, either $T(\wQ_{\ipw})$, $T(\wQ_{\conv})$ or $T(\wQ_{\aipw})$,   obtained  for the $j-$th replication. Then, we   define 
 $$L^{1,0}= \frac{1}{1000} \sum_{j=1}^{1000} |\wtheta_{j, \wpe, s }-\wtheta_{j,0}| \qquad \qquad L^{2,0}= \frac{1}{1000} \sum_{j=1}^{1000} (\wtheta_{j, \wpe, s }-\wtheta_{j,0}) ^2\,.$$ 
 We also introduce the following measures to evaluate only the effect of the missingness on the estimation procedures and the advantage of the given methods
 $$L^{1}= \frac{1}{1000} \sum_{j=1}^{1000} |\wtheta_{j, \wpe, s }-\wtheta_{j,s}|\qquad \qquad L^{2}= \frac{1}{1000} \sum_{j=1}^{1000} (\wtheta_{j, \wpe, s }-\wtheta_{j,s})^2\,,$$
 where $\wtheta_{j, s}$ stands for the estimate obtained in the $j-$replication under $C_s$ when all the data are available. These last two measures evaluate  just the effect of the missingness on the estimate, while $L^{1,0}$ and $L^{2,0}$ combines the effect that outliers and missing data have on the resulting estimator when taking $s=1$.


The obtained results are summarized in Tables \ref{tab:marg-m3} to \ref{tab:marg-m3malregre.l1l2}. More precisely, Tables \ref{tab:marg-m3} and \ref{tab:marg-m3-malregre} report bias and mean square error under $C_0$ and $C_1$ under the true nonlinear model and when the model is misspecified and fitted as a linear one, respectively. On the other hand, Tables \ref{tab:marg-m3.l1l2} and \ref{tab:marg-m3malregre.l1l2} report the new summary measures   $L^{1,0}$, $L^{2,0}$, $L^1$ and $L^2$ under $C_0$ and $C_1$ when the true model is fitted and under misspecification, respectively.


\begin{table}[H]
\caption{\label{tab:marg-m3} Summary measures for the marginal parameters under $\itM(H)$ and $\itM(1)$, for    the nonlinear model (\ref{expo}). The last block of rows denoted  $\wpe=\wpe_{c}$ corresponds to misspecification on the propensity} 
\vskip0.1in
\begin{center}
\small \begin{tabular}{c|c|rrr|rrr|}
\hline
&&\multicolumn{3}{c|}{$C_0$}&\multicolumn{3}{c|}{$C_1$}\\
\hline
$T(F)$&& Bias & sd & MSE & Bias & sd & MSE \\ 
\hline
&&\multicolumn{6}{c|}{$\itM(1)$}\\
\hline
Mean &   & -0.011 & 0.909 & 0.827 & 1.599 & 1.038 & 3.636 \\ 
 \hline
Median &   & 0.018 & 1.306 & 1.707 & 1.001 & 1.370 & 2.881 \\ 
 \hline
$M-$est &  & -0.075 & 1.158 & 1.347 & 0.674 & 1.193 & 1.879 \\ 
 \hline
&&\multicolumn{6}{c|}{$\itM(H)$, $\wpe=p$}\\
\hline
Mean & $\ipw$ & 0.035 & 1.079 & 1.166 & 1.652 & 1.235 & 4.253 \\ 
Mean & $\conv$ & 0.035 & 1.079 & 1.166 & 1.651 & 1.233 & 4.244 \\ 
Mean & $\aipw$ & -0.008 & 0.914 & 0.836 & 1.607 & 1.062 & 3.709 \\ 
\hdashline
Median & $\ipw$ & 0.121 & 1.581 & 2.514 & 1.099 & 1.670 & 3.996 \\ 
Median & $\conv$ & 0.107 & 1.556 & 2.432 & 1.487 & 1.683 & 5.044 \\ 
Median & $\aipw$& 0.033 & 1.335 & 1.784 & 1.014 & 1.423 & 3.055 \\ 
\hdashline
$M-$est & $\ipw$  & -0.045 & 1.377 & 1.898 & 0.723 & 1.412 & 2.516 \\ 
$M-$est & $\conv$  & -0.041 & 1.372 & 1.885 & 1.317 & 1.408 & 3.716 \\ 
$M-$est & $\aipw$  & -0.075 & 1.162 & 1.357 & 0.677 & 1.210 & 1.922 \\ 
 \hline
&&\multicolumn{6}{c|}{$\itM(H)$, $\wpe=\wpe_{\log}$}\\
\hline
Mean & $\ipw$ & -0.009 & 0.915 & 0.838 & 1.605 & 1.064 & 3.709 \\ 
Mean & $\conv$ & -0.009 & 0.915 & 0.838 & 1.604 & 1.061 & 3.699 \\ 
Mean & $\aipw$ & -0.009 & 0.913 & 0.833 & 1.605 & 1.060 & 3.702 \\ 
\hdashline
Median & $\ipw$ & 0.038 & 1.371 & 1.882 & 1.020 & 1.458 & 3.166 \\ 
Median & $\conv$ & 0.023 & 1.323 & 1.752 & 1.415 & 1.450 & 4.107 \\ 
Median & $\aipw$ & 0.026 & 1.330 & 1.770 & 1.012 & 1.422 & 3.045 \\ 
\hdashline
$M-$est & $\ipw$ & -0.084 & 1.178 & 1.394 & 0.673 & 1.217 & 1.933 \\ 
$M-$est & $\conv$ & -0.079 & 1.171 & 1.378 & 1.279 & 1.215 & 3.113 \\ 
$M-$est & $\aipw$ & -0.076 & 1.161 & 1.354 & 0.676 & 1.209 & 1.917 \\ 
\hline
&&\multicolumn{6}{c|}{$\itM(H)$, $\wpe=\wpe_{K}$}\\
\hline
Mean & $\ipw$ & 0.268 & 0.941 & 0.956 & 1.910 & 1.091 & 4.838 \\ 
Mean & $\conv$ & 0.268 & 0.940 & 0.956 & 1.908 & 1.089 & 4.828 \\ 
Mean & $\aipw$& 0.003 & 0.913 & 0.833 & 1.620 & 1.061 & 3.748 \\ 
\hdashline
Median & $\ipw$  & 0.427 & 1.385 & 2.100 & 1.383 & 1.482 & 4.109 \\ 
Median & $\conv$ & 0.400 & 1.344 & 1.967 & 1.799 & 1.468 & 5.390 \\ 
Median & $\aipw$ & 0.033 & 1.337 & 1.788 & 1.017 & 1.416 & 3.039 \\ 
\hdashline
$M-$est & $\ipw$ & 0.263 & 1.169 & 1.435 & 1.017 & 1.228 & 2.543 \\ 
$M-$est & $\conv$ & 0.265 & 1.166 & 1.431 & 1.608 & 1.219 & 4.072 \\ 
$M-$est & $\aipw$ & -0.067 & 1.162 & 1.356 & 0.686 & 1.209 & 1.932 \\ 
\hline
&&\multicolumn{6}{c|}{$\itM(H)$, $\wpe=\wpe_{c}$}\\
\hline
Mean & $\ipw$ & 1.150 & 1.099 & 2.532 & 2.882 & 1.277 & 9.934 \\ 
Mean & $\conv$ & 1.150 & 1.099 & 2.532 & 2.882 & 1.277 & 9.934 \\ 
Mean & $\aipw$ & 0.042 & 0.913 & 0.835 & 1.662 & 1.062 & 3.890 \\ 
\hdashline
Median & $\ipw$   & 1.725 & 1.708 & 5.894 & 2.721 & 1.792 & 10.614 \\ 
Median & $\conv$  & 1.691 & 1.676 & 5.669 & 3.111 & 1.770 & 12.811 \\ 
Median & $\aipw$  & 0.061 & 1.333 & 1.781 & 1.061 & 1.422 & 3.148 \\ 
\hdashline
$M-$est & $\ipw$ & 1.318 & 1.307 & 3.446 & 2.105 & 1.400 & 6.389 \\ 
$M-$est & $\conv$ & 1.321 & 1.305 & 3.448 & 2.639 & 1.368 & 8.836 \\ 
$M-$est & $\aipw$ & -0.034 & 1.164 & 1.357 & 0.722 & 1.211 & 1.989 \\ 
 \hline
\end{tabular}
\end{center}
\end{table}


\begin{table}[ht!]
\caption{\label{tab:marg-m3-malregre} Summary measures for the marginal parameters under $\itM(H)$ and    the nonlinear model (\ref{expo}), when the regression model is misspecified.} 
\vskip0.1in
\begin{center}
\small 
\begin{tabular}{c|c|rrr|rrr|}
\hline
&&\multicolumn{3}{c|}{$C_0$}&\multicolumn{3}{c|}{$C_1$}\\
\hline
$T(F)$&& Bias & sd & MSE & Bias & sd & MSE \\ 
\hline
&&\multicolumn{6}{c|}{$\wpe=p$}\\
\hline
Mean & $\ipw$ & 0.035 & 1.079 & 1.166 & 1.652 & 1.235 & 4.253 \\ 
Mean & $\conv$ & -0.037 & 1.079 & 1.166 & 1.573 & 1.231 & 3.991 \\ 
Mean & $\aipw$ & -0.008 & 0.914 & 0.836 & 1.607 & 1.062 & 3.709 \\ 
\hdashline
Median & $\ipw$ & 0.121 & 1.581 & 2.514 & 1.099 & 1.670 & 3.996 \\ 
Median & $\conv$ & 2.285 & 1.609 & 7.812 & 3.548 & 1.652 & 15.317 \\ 
Median & $\aipw$ & 0.033 & 1.335 & 1.784 & 1.014 & 1.423 & 3.055 \\ 
\hdashline
$M-$est & $\ipw$ & -0.045 & 1.377 & 1.898 & 0.723 & 1.412 & 2.516 \\ 
$M-$est & $\conv$ & 0.605 & 1.149 & 1.687 & 1.773 & 1.244 & 4.690 \\ 
$M-$est & $\aipw$ & -0.075 & 1.162 & 1.357 & 0.677 & 1.210 & 1.922 \\ 
\hline
&&\multicolumn{6}{c|}{$\wpe=\wpe_{\log}$}\\
\hline
Mean & $\ipw$ & -0.009 & 0.915 & 0.838 & 1.605 & 1.064 & 3.709 \\ 
Mean & $\conv$ & -0.101 & 0.914 & 0.846 & 1.504 & 1.058 & 3.381 \\ 
Mean & $\aipw$ & -0.009 & 0.913 & 0.833 & 1.605 & 1.060 & 3.702 \\ 
\hdashline
Median & $\ipw$ & 0.038 & 1.371 & 1.882 & 1.020 & 1.458 & 3.166 \\ 
Median & $\conv$ & 2.200 & 1.379 & 6.740 & 3.481 & 1.430 & 14.164 \\ 
Median & $\aipw$ & 0.026 & 1.330 & 1.770 & 1.012 & 1.422 & 3.045 \\ 
\hdashline
$M-$est & $\ipw$ & -0.084 & 1.178 & 1.394 & 0.673 & 1.217 & 1.933 \\ 
$M-$est & $\conv$ & 0.533 & 0.969 & 1.223 & 1.711 & 1.065 & 4.062 \\ 
$M-$est & $\aipw$ & -0.076 & 1.161 & 1.354 & 0.676 & 1.209 & 1.917 \\ 
\hline
&&\multicolumn{6}{c|}{$\wpe=\wpe_{K}$}\\
\hline
Mean & $\ipw$ & 0.268 & 0.941 & 0.956 & 1.910 & 1.091 & 4.838 \\ 
Mean & $\conv$ & 0.217 & 0.940 & 0.931 & 1.854 & 1.088 & 4.623 \\ 
Mean & $\aipw$ & 0.003 & 0.913 & 0.833 & 1.620 & 1.061 & 3.748 \\ 
\hdashline
Median & $\ipw$ & 0.427 & 1.385 & 2.100 & 1.383 & 1.482 & 4.109 \\ 
Median & $\conv$ & 2.621 & 1.386 & 8.788 & 3.877 & 1.433 & 17.084 \\ 
Median & $\aipw$ & 0.033 & 1.337 & 1.788 & 1.017 & 1.416 & 3.039 \\ 
\hdashline
$M-$est & $\ipw$ & 0.263 & 1.169 & 1.435 & 1.017 & 1.228 & 2.543 \\ 
$M-$est & $\conv$ & 0.870 & 0.998 & 1.753 & 2.031 & 1.088 & 5.309 \\ 
$M-$est & $\aipw$& -0.067 & 1.162 & 1.356 & 0.686 & 1.209 & 1.932 \\ 
\hline
 
\end{tabular}
\end{center}

\end{table}
We first summarize the results  under $C_0$ in terms of the classical measures, i.e.,  bias, standard deviation and mean square error. As shown in Table \ref{tab:marg-m3},  the bias of the   augmented inverse probability weighting estimator is smaller than that of  $\wtheta_{\ipw}$ and $\wtheta_{\conv}$,  when the propensity model is estimated using the correct model or with kernels. The only exception corresponds to the median that is estimated with a smaller biased when the correct missing probability model is specified and the convolution--based method is used. As expected, under propensity misspecification, the bias of both $\wtheta_{\ipw}$ and $\wtheta_{\conv}$ are enlarged, while $\wtheta_{\aipw}$ still leads to reliable bias results. It should be noted that even when using a kernel approach to estimate the propensity, the \textsc{aipw} procedure leads to smaller biases and standard deviations than the \textsc{ipw} method, which, in this case, also provides consistent estimators.

\begin{figure}[ht!]

\begin{center}
\setlength{\tabcolsep}{0.1pt} 
\renewcommand{\arraystretch}{0.6}
\begin{tabular}{cc  }
  \small $C_0$ & $C_1$   \\
 
     \includegraphics[scale=0.4]{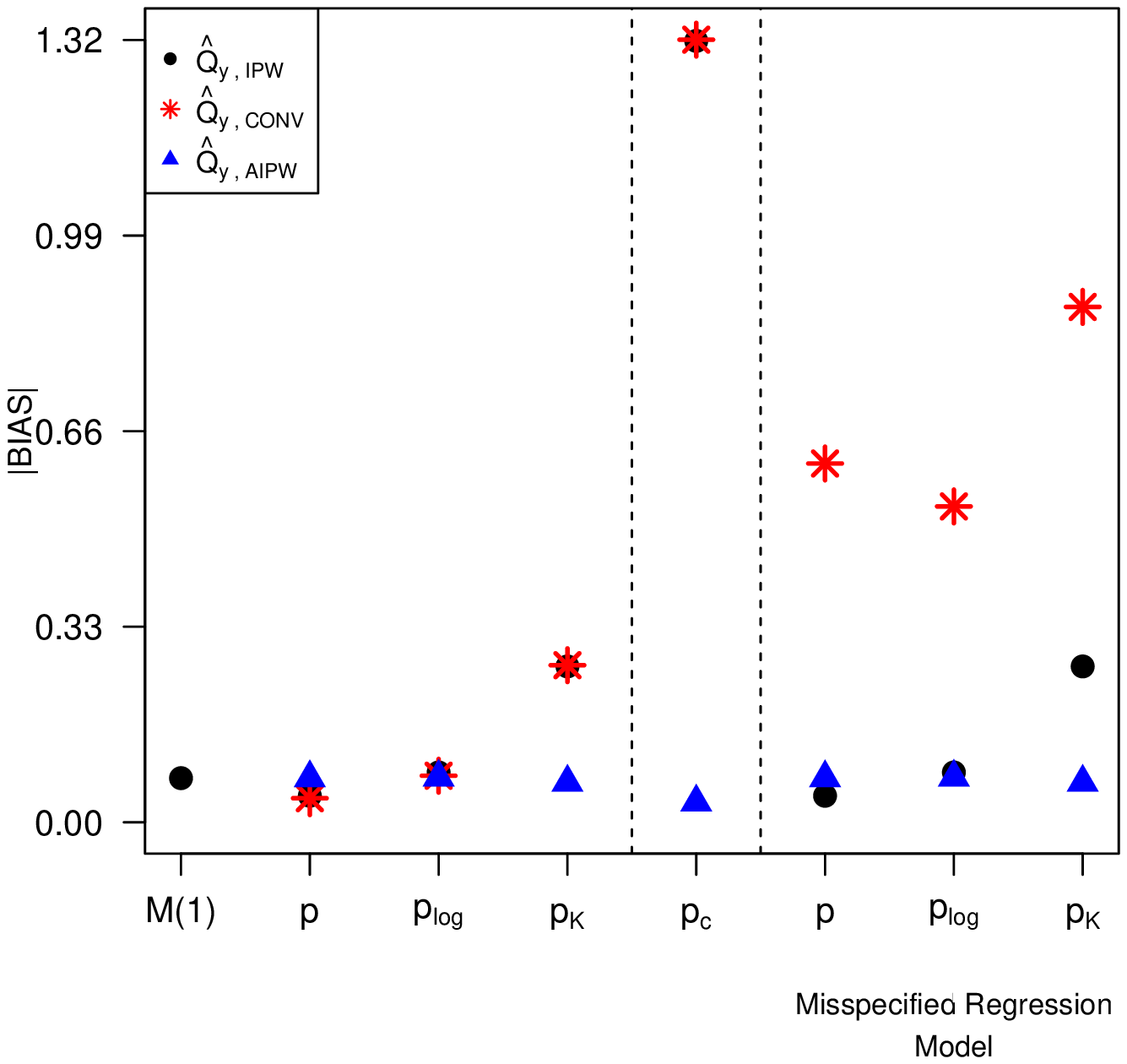}   &
     \includegraphics[scale=0.4]{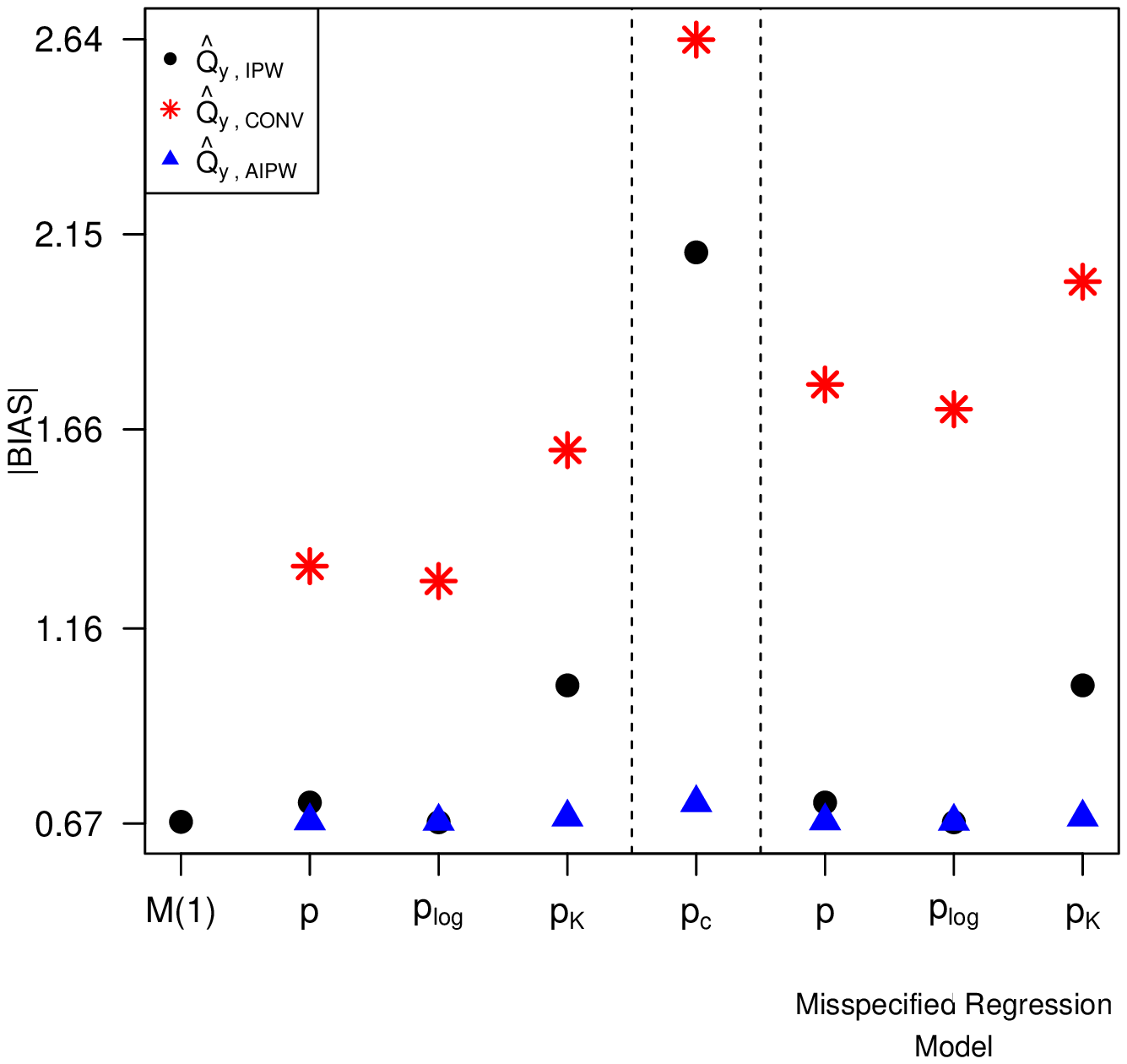}   
      
    \\
     
     \includegraphics[scale=0.4]{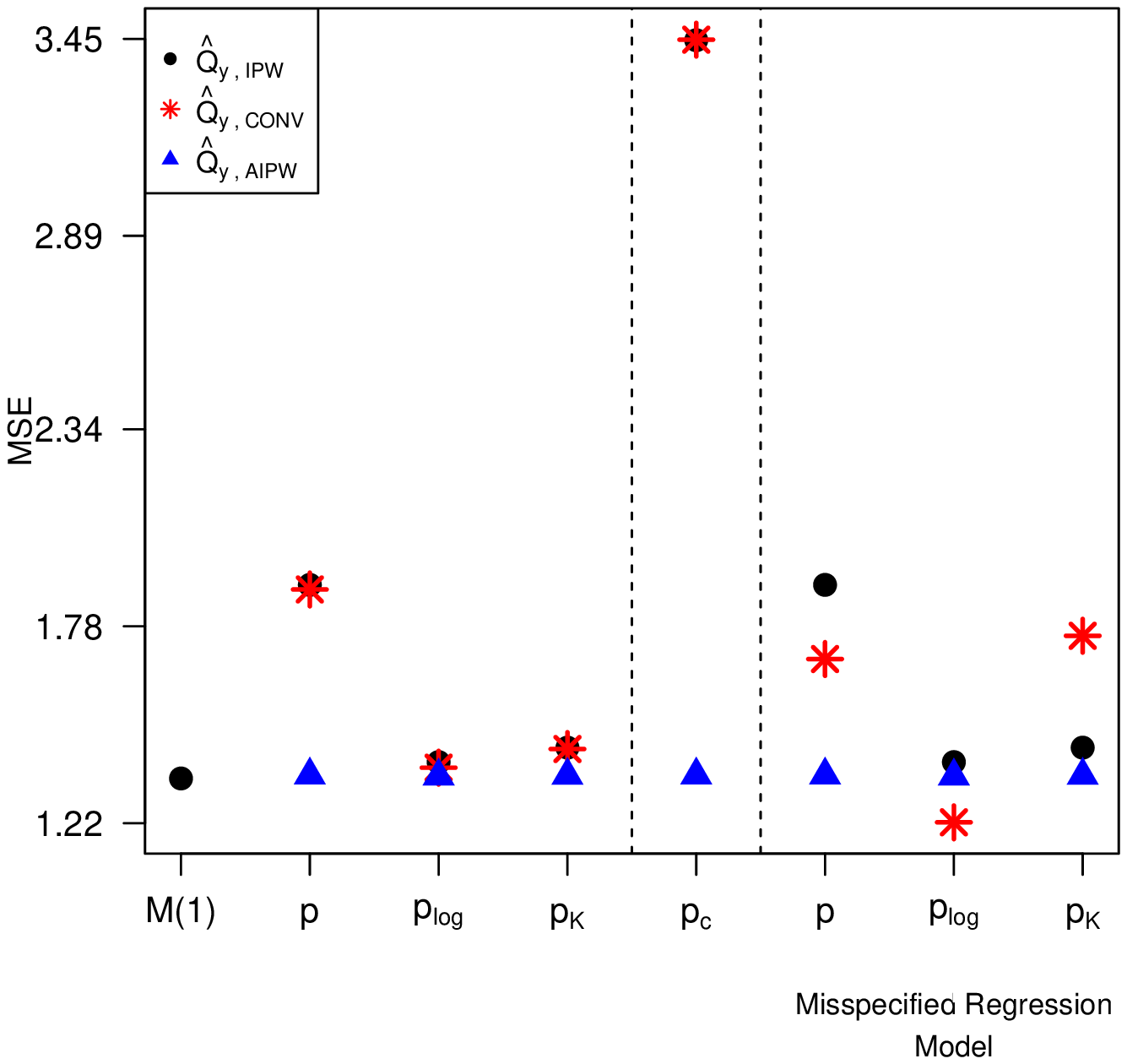}   &
     \includegraphics[scale=0.4]{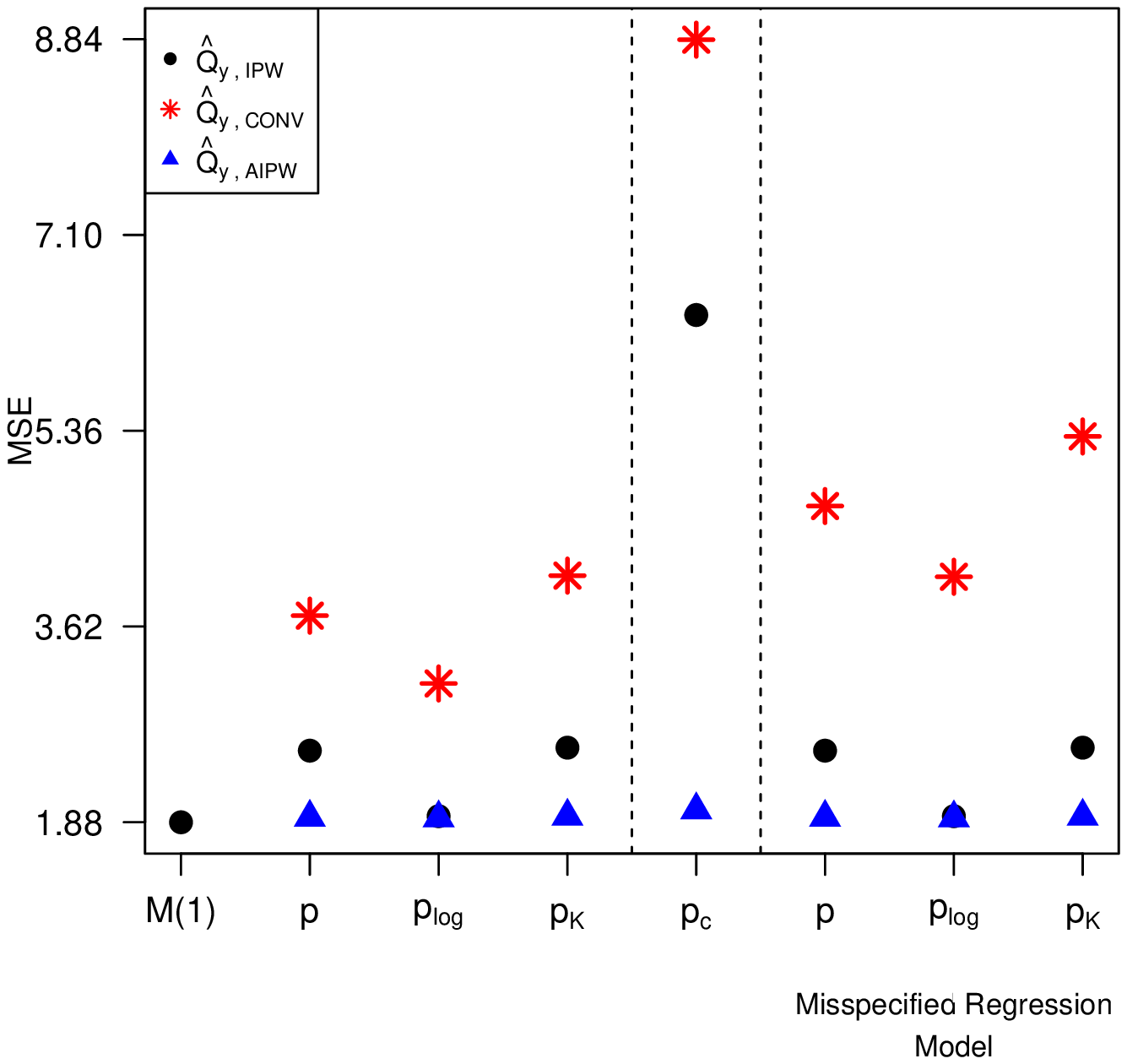}

\end{tabular}
\vskip0.1in
 \caption{\label{fig:Bias-Mest-NL} Bias (upper plots) and MSE (lower plots) of the $M-$location estimator under $C_0$ and $C_1$ for the nonlinear model (\ref{expo}). The summary measures for $\wtheta=T_\rho(\wQ_y)$ for $\wQ_y=\wQ_{y,\ipw}$, $ \wQ_{y,\conv}$ and $ \wQ_{y,\aipw}$ are given in black dotted points, red stars and   blue filled triangles, respectively.}
\end{center}
\end{figure}

In Table \ref{tab:marg-m3}  we also observe that the \textsc{aipw} procedure always results in more efficient estimators. The only exception corresponds to the median when the estimation procedure is the convolution--based method and  both, the regression and propensity models, are correctly specified. The mean square error of the estimators based on the \textsc{aipw}  are the smallest, except for the referred case of the median. The bias and the mean square errors of  the $M-$estimators are plotted in Figure \ref{fig:Bias-Mest-NL}. The black dotted points correspond to the summary measures of $\wtheta_{\ipw}$, the red stars to those of $\wtheta_{\conv}$, while the blue triangles indicate the results obtained when using the \textsc{aipw} procedure. In particular, the left panels of Figure \ref{fig:Bias-Mest-NL} show the great impact of propensity misspecification on the estimators obtained with the convolution--based method. At the same time, this figure reveals the gain in bias and MSE of the $M-$estimators based on the augmented inverse probability weighting method. Table \ref{tab:marg-m3-malregre}, where the results under misspecification of the regression model are exhibited, shows that when the propensity is estimated with the right model or with kernels, the bias and the MSE of the estimators computed with the \textsc{aipw} approach are the smallest ones.


If we restrict our comparison to the \textsc{ipw} and convolution--based methods, Table \ref{tab:marg-m3} also shows that  in most situations either for the mean, the median or the $M-$estimator, the standard deviations obtained with the convolution--based method are smaller or equal than those obtained with the inverse probability weighting procedure, when the regression model is correct and no matter if the propensity is estimated or not and  if its estimation  is  based on a correct model or on a misspecified one. The same assertion can be made with respect to mean square error of the three location measures considered, except for the case of the $M-$estimator when the propensity is misspecified, due to an increase of the bias of the estimator based on the convolution method.

Under $C_1$, the results go in the same direction. Indeed, from Table  \ref{tab:marg-m3}, we can  conclude that the MSE and the standard deviations of the estimators based on \textsc{aipw} method are the smallest ones. Regarding the bias of the $M-$estimators, the \textsc{aipw} estimators  outperform the other two procedures when the propensity is estimated through kernels or even if the missing probability  model is misspecified. Besides, the right panels of Figure \ref{fig:Bias-Mest-NL} illustrate the greater stability of the $M-$estimators based on the \textsc{aipw} method either in terms of bias or  MSE when comparing them with the \textsc{ipw} and convolution--based methods. Furthermore, when the regression model is misspecified, as reported in Table \ref{tab:marg-m3-malregre}, the \textsc{aipw} method leads to estimators with the lowest MSE values,   as expected.
 
Now, focusing on the new measures $L^{2}$ and $L^{20}$,  Tables \ref{tab:marg-m3.l1l2} and \ref{tab:marg-m3malregre.l1l2} show that, under $C_0$ and $C_1$, in the majority of the analysed situations, the estimators based on the augmented inverse probability weighting method achieve the lowest   $L^{2}$  values. This is still valid when only one of the models, i.e., the regression  or the propensity model, is correctly specified. When the effect on the estimators of both the contamination and the missingness is measured through $L^{1}$ and $L^{10}$, the conclusions are almost similar.

In conclusion, regarding the  performance of the $M-$location estimators,    the augmented inverse probability weighting procedure improves the performance   of the estimators. More precisely,  under the true regression and propensity models,   $\wtheta_{\aipw}$ outperforms $\wtheta_{\ipw}$ and $\wtheta_{\conv}$ in bias and   the mean square error,  for clean and contaminated samples. The same assertion holds under misspecification of the propensity or the regression model. Only a smaller mean square error   has been observed in our simulation study for   $\wtheta_{\conv}$ when the regression is incorrectly specified and the propensity is estimated under the true logistic model, even though the bias is very large (see Figure \ref{fig:Bias-Mest-NL}). This reduction can be explained by a decrease of around 10\% in the standard deviation. However, the inverse probability weighting and the convolution based estimators show their weakness to estimate the marginal location, when one of the models is misspecified. For all  these reasons and taking into account the stability of the $M-$estimator under contamination, it is better to bet on the augmented inverse probability weighted $M-$estimators which protect against the considered deviations from the underlying models. 



\begin{table}[H]
\caption{\label{tab:marg-m3.l1l2} New summary measures for the marginal parameters under $\itM(H)$ and    the nonlinear model (\ref{expo}). The last block of rows denoted  $\wpe=\wpe_{c}$ corresponds to misspecification on the propensity.} 
\vskip0.1in
\begin{center}
\small 
\begin{tabular}{c|c|rr|rr|rr|rr|}
\hline
&&\multicolumn{2}{c|}{$C_0$}&\multicolumn{2}{c|}{$C_1$}&\multicolumn{2}{c|}{$C_0$}&\multicolumn{2}{c|}{$C_1$}\\
\hline
$T(F)$&$\wpe$&$L^{1,0}$ & $L^{2,0}$ & $L^{1,0}$ & $L^{2,0}$ & $L^{1}$ &  $L^{2}$ &   $L^{1}$ &  $L^{2}$\\ 
\hline
&&\multicolumn{8}{c|}{$\wpe=p$}\\
\hline
Mean & $\ipw$ & 0.410 & 0.277 & 1.665 & 3.243 & 0.410 & 0.277 & 0.497 & 0.398 \\ 
Mean & $\conv$  & 0.410 & 0.277 & 1.664 & 3.237 & 0.410 & 0.277 & 0.501 & 0.401 \\ 
Mean & $\aipw$ & 0.057 & 0.005 & 1.618 & 2.767 & 0.057 & 0.005 & 0.207 & 0.070 \\ 
\hdashline
Median & $\ipw$ & 0.646 & 0.822 & 1.192 & 2.415 & 0.646 & 0.822 & 0.700 & 0.954 \\ 
Median & $\conv$  & 0.704 & 0.849 & 1.513 & 3.291 & 0.704 & 0.849 & 0.905 & 1.396 \\ 
Median & $\aipw$ & 0.215 & 0.095 & 1.007 & 1.503 & 0.215 & 0.095 & 0.323 & 0.231 \\ 
\hdashline
$M-$est & $\ipw$ & 0.522 & 0.454 & 0.941 & 1.297 & 0.522 & 0.454 & 0.565 & 0.512 \\ 
$M-$est & $\conv$  & 0.525 & 0.455 & 1.409 & 2.493 & 0.525 & 0.455 & 0.845 & 1.127 \\ 
$M-$est & $\aipw$ & 0.063 & 0.007 & 0.777 & 0.783 & 0.063 & 0.007 & 0.199 & 0.065 \\ 
\hline
&&\multicolumn{8}{c|}{$\wpe=\wpe_{\log}$}\\
\hline
Mean & $\ipw$ & 0.067 & 0.007 & 1.616 & 2.764 & 0.067 & 0.007 & 0.211 & 0.071 \\ 
Mean & $\conv$  & 0.065 & 0.007 & 1.616 & 2.759 & 0.065 & 0.007 & 0.217 & 0.075 \\ 
Mean & $\aipw$ & 0.055 & 0.005 & 1.617 & 2.762 & 0.055 & 0.005 & 0.207 & 0.070 \\ 
\hdashline
Median & $\ipw$ & 0.314 & 0.211 & 1.022 & 1.573 & 0.314 & 0.211 & 0.369 & 0.298 \\ 
Median & $\conv$  & 0.384 & 0.237 & 1.400 & 2.424 & 0.384 & 0.237 & 0.642 & 0.693 \\ 
Median & $\aipw$& 0.215 & 0.094 & 1.005 & 1.494 & 0.215 & 0.094 & 0.320 & 0.224 \\ 
\hdashline
$M-$est & $\ipw$ & 0.109 & 0.024 & 0.774 & 0.780 & 0.109 & 0.024 & 0.209 & 0.071 \\ 
$M-$est & $\conv$  & 0.122 & 0.028 & 1.354 & 1.966 & 0.122 & 0.028 & 0.665 & 0.664 \\ 
$M-$est & $\aipw$ & 0.060 & 0.006 & 0.776 & 0.780 & 0.060 & 0.006 & 0.198 & 0.064 \\ 

\hline
&&\multicolumn{8}{c|}{$\wpe=\wpe_{K}$}\\
\hline
Mean & $\ipw$ & 0.290 & 0.113 & 1.922 & 3.882 & 0.290 & 0.113 & 0.376 & 0.197 \\ 
Mean & $\conv$  & 0.290 & 0.113 & 1.920 & 3.874 & 0.290 & 0.113 & 0.376 & 0.199 \\ 
Mean & $\aipw$ & 0.056 & 0.005 & 1.631 & 2.809 & 0.056 & 0.005 & 0.209 & 0.070 \\ 
\hdashline
Median & $\ipw$ & 0.455 & 0.394 & 1.367 & 2.536 & 0.455 & 0.394 & 0.488 & 0.476 \\ 
Median & $\conv$  & 0.497 & 0.386 & 1.782 & 3.682 & 0.497 & 0.386 & 0.876 & 1.178 \\ 
Median & $\aipw$& 0.213 & 0.092 & 1.010 & 1.513 & 0.213 & 0.092 & 0.325 & 0.237 \\ 
\hdashline
$M-$est & $\ipw$ & 0.348 & 0.163 & 1.100 & 1.457 & 0.348 & 0.163 & 0.388 & 0.219 \\ 
$M-$est & $\conv$  & 0.353 & 0.169 & 1.683 & 2.993 & 0.353 & 0.169 & 0.948 & 1.188 \\ 
$M-$est & $\aipw$ & 0.060 & 0.006 & 0.786 & 0.799 & 0.060 & 0.006 & 0.200 & 0.066 \\ 

\hline
&&\multicolumn{8}{c|}{$\wpe=\wpe_{c}$}\\
\hline
Mean & $\ipw$ & 1.167 & 1.622 & 2.893 & 8.899 & 1.167 & 1.622 & 1.294 & 2.052 \\ 
Mean & $\conv$  & 1.167 & 1.622 & 2.893 & 8.899 & 1.167 & 1.622 & 1.294 & 2.052 \\ 
Mean & $\aipw$ & 0.072 & 0.008 & 1.673 & 2.951 & 0.072 & 0.008 & 0.220 & 0.074 \\ 
\hdashline
Median & $\ipw$ & 1.717 & 4.149 & 2.704 & 8.991 & 1.717 & 4.149 & 1.729 & 4.307 \\ 
Median & $\conv$  & 1.692 & 3.931 & 3.094 & 11.082 & 1.692 & 3.931 & 2.119 & 5.941 \\ 
Median & $\aipw$ & 0.219 & 0.098 & 1.054 & 1.619 & 0.219 & 0.098 & 0.334 & 0.243 \\ 
\hdashline
$M-$est & $\ipw$ & 1.400 & 2.342 & 2.181 & 5.406 & 1.400 & 2.342 & 1.438 & 2.553 \\ 
$M-$est & $\conv$  & 1.402 & 2.352 & 2.714 & 7.902 & 1.402 & 2.352 & 1.968 & 4.557 \\ 
$M-$est & $\aipw$& 0.071 & 0.008 & 0.819 & 0.858 & 0.071 & 0.008 & 0.207 & 0.069 \\ 
\hline
\end{tabular}
\end{center}
\end{table}

\newpage

\begin{table}[ht!]

\caption{\label{tab:marg-m3malregre.l1l2} New summary measures for the marginal parameters under $\itM(H)$ and    the nonlinear model (\ref{expo}),  when the regression model is misspecified.} 
\vskip0.1in
\begin{center}
\small 
\begin{tabular}{c|c|rr|rr|rr|rr|}
\hline
&&\multicolumn{2}{c|}{$C_0$}&\multicolumn{2}{c|}{$C_1$}&\multicolumn{2}{c|}{$C_0$}&\multicolumn{2}{c|}{$C_1$}\\
\hline
$T(F)$&$\wpe$&$L^{1,0}$ & $L^{2,0}$ & $L^{1,0}$ & $L^{2,0}$ & $L^{1}$ &  $L^{2}$ &   $L^{1}$ &  $L^{2}$\\ 
\hline
&&\multicolumn{8}{c|}{$\wpe=p$}\\
\hline
Mean & $\ipw$ & 0.410 & 0.277 & 1.665 & 3.243 & 0.410 & 0.277 & 0.497 & 0.398 \\ 
Mean & $\conv$  & 0.412 & 0.277 & 1.587 & 2.982 & 0.412 & 0.277 & 0.498 & 0.397 \\ 
Mean & $\aipw$ & 0.057 & 0.005 & 1.618 & 2.767 & 0.057 & 0.005 & 0.207 & 0.070 \\ 
\hdashline
Median & $\ipw$ & 0.646 & 0.822 & 1.192 & 2.415 & 0.646 & 0.822 & 0.700 & 0.954 \\ 
Median & $\conv$  & 2.270 & 6.064 & 3.530 & 13.577 & 2.270 & 6.064 & 2.548 & 7.675 \\ 
Median & $\aipw$ & 0.215 & 0.095 & 1.007 & 1.503 & 0.215 & 0.095 & 0.323 & 0.231 \\ 
\hdashline
$M-$est & $\ipw$ & 0.522 & 0.454 & 0.941 & 1.297 & 0.522 & 0.454 & 0.565 & 0.512 \\ 
$M-$est & $\conv$  & 0.758 & 0.843 & 1.849 & 3.899 & 0.758 & 0.843 & 1.140 & 1.789 \\ 
$M-$est & $\aipw$ & 0.063 & 0.007 & 0.777 & 0.783 & 0.063 & 0.007 & 0.199 & 0.065 \\ 
\hline
&&\multicolumn{8}{c|}{$\wpe=\wpe_{\log}$}\\
\hline
Mean & $\ipw$ & 0.067 & 0.007 & 1.616 & 2.764 & 0.067 & 0.007 & 0.211 & 0.071 \\ 
Mean & $\conv$  & 0.115 & 0.021 & 1.515 & 2.447 & 0.115 & 0.021 & 0.229 & 0.087 \\ 
Mean & $\aipw$ & 0.055 & 0.005 & 1.617 & 2.762 & 0.055 & 0.005 & 0.207 & 0.070 \\ 
\hdashline
Median & $\ipw$ & 0.314 & 0.211 & 1.022 & 1.573 & 0.314 & 0.211 & 0.369 & 0.298 \\ 
Median & $\conv$  & 2.182 & 5.079 & 3.463 & 12.494 & 2.182 & 5.079 & 2.480 & 6.750 \\ 
Median & $\aipw$ & 0.215 & 0.094 & 1.005 & 1.494 & 0.215 & 0.094 & 0.320 & 0.224 \\ 
\hdashline
$M-$est & $\ipw$ & 0.109 & 0.024 & 0.774 & 0.780 & 0.109 & 0.024 & 0.209 & 0.071 \\ 
$M-$est & $\conv$  & 0.608 & 0.458 & 1.786 & 3.340 & 0.608 & 0.458 & 1.039 & 1.331 \\ 
$M-$est & $\aipw$ & 0.060 & 0.006 & 0.776 & 0.780 & 0.060 & 0.006 & 0.198 & 0.064 \\ 
\hline
&&\multicolumn{8}{c|}{$\wpe=\wpe_{K}$}\\
\hline
Mean & $\ipw$ & 0.290 & 0.113 & 1.922 & 3.882 & 0.290 & 0.113 & 0.376 & 0.197 \\ 
Mean & $\conv$  & 0.261 & 0.096 & 1.866 & 3.677 & 0.261 & 0.096 & 0.349 & 0.175 \\ 
Mean & $\aipw$ & 0.056 & 0.005 & 1.631 & 2.809 & 0.056 & 0.005 & 0.209 & 0.070 \\ 
\hdashline
Median & $\ipw$& 0.455 & 0.394 & 1.367 & 2.536 & 0.455 & 0.394 & 0.488 & 0.476 \\ 
Median & $\conv$  & 2.603 & 7.117 & 3.859 & 15.434 & 2.603 & 7.117 & 2.876 & 8.906 \\ 
Median & $\aipw$ & 0.213 & 0.092 & 1.010 & 1.513 & 0.213 & 0.092 & 0.325 & 0.237 \\ 
\hdashline
$M-$est & $\ipw$ & 0.348 & 0.163 & 1.100 & 1.457 & 0.348 & 0.163 & 0.388 & 0.219 \\ 
$M-$est & $\conv$ & 0.944 & 1.016 & 2.106 & 4.630 & 0.944 & 1.016 & 1.358 & 2.128 \\ 
$M-$est & $\aipw$ & 0.060 & 0.006 & 0.786 & 0.799 & 0.060 & 0.006 & 0.200 & 0.066 \\ 
\hline
 
\end{tabular}
\end{center}

\end{table}

\section{Ozone concentration Data}{\label{sec:ejemplo}}

In Cleveland (1985) a data set of 153 daily measurements of ozone (ppb) and wind speed (mph) is considered. The data were collected in New York metropolitan area between May 1, 1973 and September 30, 1973.
Cleveland (1985) finds out a decreasing nonlinear relationship between ozone and wind speed  that explains the ventilation that is produced by higher wind speeds.
In our study, we also include as linear component a third variable that records the solar radiation. It is worth noticing that even when all the values of wind speed are present,  37 observations of ozone and 7 values of solar radiation are dropped out. 

Bianco and Spano (2017) fit an exponential growth model for variable ozone using  wind speed as independent variable. For this purpose, those authors implement a weighted $MM-$estimator and their analysis enables the identification of five outliers (corresponding to observations labelled as 86, 100, 101, 121 and 126). Taking into account the well known sensitivity of the mean to the presence to anomalous data, henceforth we focus on a marginal $M-$location parameter. The five atypical observations mentioned above are kept in our analysis in order to challenge the robust marginal estimator.

Table \ref{tab:ejemplo}  summarizes the obtained  $M-$estimators of the marginal distribution based on the inverse probability method, the convolution--based estimator and the augmented inverse probability procedure,  i.e., $\wtheta_{\ipw}$, $\wtheta_{\conv}$ and $\wtheta_{\aipw}$, respectively. 
Each of them is computed from a constant propensity $p_{c}$, a logistic propensity  $p_{\log}$ and  using a nonparametric approach based on a kernel  estimator using the Epanechnikov function, $p_{K}$. As in our simulation study, the marginal $M-$location $\wtheta=T_\rho(\wQ_y)$ uses as $\rho-$function the bisquare function with tuning constant 4.685 and as preliminary scale estimator   an $S-$estimator with $b=0.5$. 

The inverse probability and augmented inverse probability method estimators do not depend on a regression fit, while the convolution--based estimator does. To calculate the predicted values that are needed for $\wtheta_{\conv}$, we consider two models. In a first stage, we propose a similar nonlinear model to that given in equation \eqref{expo} with an exponential component based on wind speed, while it depends linearly on solar radiation, that is,
\begin{equation}
y=\beta_1 \exp({\beta_2 x_1})+\beta_3 \, + \beta_4 x_2 + \epsilon \, , \label{ejemplo1}
\end{equation}
while in a second stage, we fit a linear model based on both covariates given by
\begin{equation}
y=\wtbeta_1 x_1+\wtbeta_2 x_2 + \widetilde{\beta}_3+\epsilon \, , \label{ejemplo2}
\end{equation}
where $y$, $x_1$ and $x_2$ represent the variables ozone, wind speed and solar radiation, respectively. Hence, in this case $\bz_i=x_{i1}$ and $\bz_i^{(m)}=(y_i,x_{i2})$, with $\bx_i=(x_{i1},x_{i2})\trasp$, $1 \le i \le n=153$.
In the case of the nonlinear model, we compute a weighted simplified $MM-$estimator of the parameters with weights based on a continuous version of a hard--rejection type function applied to the covariate wind speed, while for the linear model the coefficients are fitted using a simplified $MM-$estimator. 
 

\begin{table}[ht!]

\caption{\label{tab:ejemplo}Marginal $M-$estimators.}
\vskip0.1in
\begin{center}
\renewcommand{\arraystretch}{1.2}
\setlength{\tabcolsep}{4pt}
\small\begin{tabular}{ |c| c c c|}
\hline 
$\wpe$ & $\wpe_{\log}$& $\wpe_{K}$ & $\wpe_{c}$\\
\hline 
   $\wtheta_{\ipw}$ & 35.848 & 35.805 & 35.954 \\
   $\wtheta_{\aipw}$ & 35.802 & 35.787 & 35.832 \\
\hline
    & \multicolumn{3}{c|}{Nonlinear Fit}\\
\hline
      $\wtheta_{\conv}$ & 36.051 & 36.055 & 36.126 \\
\hline
   &\multicolumn{3}{c|}{Linear Fit}\\
\hline
   $\wtheta_{\conv}$ & 41.020 & 40.992 & 41.107\\
\hline 
\end{tabular} 
\end{center}

\end{table}

As shown in Table \ref{tab:ejemplo}, $\wtheta_{\conv}$ is very sensitive to the inadequacy of the linear model fit.  Note that, under a missing at random model, $\wtheta_{\ipw}$ computed with a kernel is naturally protected against propensity misspecification. Since $\wtheta_{\aipw}$ protects against  misspecification both on the regression and the propensity  models, the similarity  between both estimators is very natural.

\begin{figure}[ht]
\begin{center}
\includegraphics[scale=0.4]{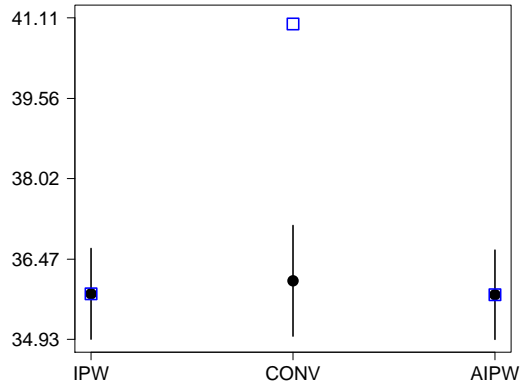}
\vskip0.1in
\caption{ \label{fig:ejemplo} Air quality data: 95\% asymptotic confidence intervals based on $\wtheta_{\ipw}$ , $\wtheta_{\conv}$  and $\wtheta_{\aipw}$ using a nonlinear regression model.  The blue squares correspond to the estimated values under the lineal model.} 
\end{center}
\end{figure}

We also compute  the jackniffe standard deviations of  $\wtheta_{\ipw}$, $\wtheta_{\conv}$  and $\wtheta_{\aipw}$ based on the propensity estimated by $p_{K}$. For the convolution--based estimator  the nonlinear model \eqref{ejemplo1} is fitted. These standard deviations are equal to   0.4446, 0.5424 and 0.4377, respectively. From these estimates,  we build 95\% asymptotic confidence intervals which are shown in  Figure \ref{fig:ejemplo}, which reveals that  the interval corresponding to $\wtheta_{\aipw}$ is the shortest. The central black dot on each interval corresponds to its center, that is the obtained estimate in each case, while the blue squares correspond to the estimated values under the lineal model. It is evident that the value of $\wtheta_{\conv}$ obtained under the linear model lies outside the interval, while  $\wtheta_{\ipw}$ and $\wtheta_{\aipw}$ are not affected by the  fitted regression model. 
 

\section{Final Remarks}{\label{sec:conclusion}}
In this paper, we introduce robust $M-$marginal location estimators when   missing data arise both in responses and on some of the covariates under a general \textsc{mar} missing scheme by  plugging--in a properly defined  marginal distribution estimator on the $M-$functional.   The  considered approach includes inverse probability weighting, convolution--based estimators and also an augmented inverse probability procedure that protects against misspecification of the regression model or the propensity scheme. The convergence of the marginal distribution estimators   allows to obtain consistent estimators  $M-$marginal location estimators. Furthermore,  the asymptotic distribution results obtained for the inverse probability weighted and the convolution based $M-$estimators allow to quantify the advantage of the last ones when both the regression and propensity models are correct. As shown in our simulation study, when estimating the mean and $M-$location parameters, the
augmented inverse probability estimators are more accurate, under a nonlinear regression model, leading to more reliable results under misspecification. 

\vskip0.1in

\noi {\small\textbf{Acknowledgment.} This work was partially developed while  Ana M. Bianco and Graciela Boente were visiting the Departamento de Estat\'\i stica, An\'alise Matem\'atica e Optimizaci\'on de la Universidad de Santiago de Compostela, Spain under the bilateral agreement between the Universidad de Buenos Aires and the Universidad de Santiago de Compostela.  This research was partially supported by  \textsc{anpcyt} in  Argentina under Grant   \textsc{pict}   2018-00740,   Universidad de Buenos Aires in  Argentina under Grant 20020170100022BA  and also by the Ministry of Economy and Competitiveness in Spain (MINECO/AEI/FEDER, UE) under the Spanish Project {MTM2016-76969P}. A. Bianco and G. Boente also wish to thank the Minerva Foundation for its support to present some of this paper results at the International Conference on Robust Statistics 2017.}  

\normalsize

\small
\section*{References}
\begin{description}

\item Bianco, A. and Boente, G. (2004). Robust estimators in semiparametric partly linear regression models. \textsl{Journal of Statistical Planning and
Inference}, \textbf{122},  229-252.

\item Bianco, A.; Boente, G.; Gonz\'alez--Manteiga, W. and P\'erez--Gonz\'alez, A. (2010). Estimation of the marginal location under a partially linear model with missing responses.  \textsl{Computational Statistics and Data Analysis}, \textbf{54}, 546-564.

\item Bianco, A.; Boente, G.; Gonz\'alez--Manteiga, W. and P\'erez--Gonz\'alez, A. (2011). Asymptotic behavior of robust estimators in partially linear models with missing responses: The effect of estimating the missing probability on the simplified marginal estimators. \textsl{TEST}, \textbf{20}, 524-548.

\item Bianco, A.; Boente, G.; Gonz\'alez--Manteiga, W. and P\'erez--Gonz\'alez, A. (2018). Plug--in marginal estimation  under a general regression model with missing responses and covariates.  In press in \textsl{TEST}. \url{https://doi.org/10.1007/s11749-018-0591-5}

\item  Bianco, A. and Spano, P. (2017). Robust inference for nonlinear regression models. In press in \textsl{TEST}. \url{ https://doi.org/10.1007/s11749-018-0591-5}

\item Boente, G.; Gonz\'alez--Manteiga, W. and P\'erez--Gonz\'alez, A. (2009). Robust nonparametric estimation with missing data. \textsl{Journal of Statistical Planning and Inference}, \textbf{139}, 571-592.

\item Cantoni, E. and Ronchetti, E. (2006). A robust approach for skewed and heavy-tailed outcomes in the analysis of health care expenditures. \textsl{Journal of Health Economics}, \textbf{25}, 198-213. 

\item
 {Chen, Q.}; {Ibrahim, J.}; {Chen, M.} and {Senchaudhuri, P.} (2008). Theory and inference for regression models with missing responses and covariates, \textsl{Journal of Multivariate Analysis}, \textbf{99}, 1302-1331.

\item
 {Chen, X.}; {Wan, A.} and {Zhou, Y.} (2015). Efficient quantile regression analysis with missing observations. \textsl{Journal of the American Statistical Association}, \textbf{110}, 723-741.

 \item Cleveland, W. (1985). \textsl{The elements of graphing data}.  Bell Telephone Laboratories Inc., New Jersey.

\item D\'{\i}az, I. (2017). Efficient estimation of quantiles in missing data models. \textsl{Journal of Statistical Planning and Inference}, \textbf{190}, 39-51.

\item Fasano, V. (2009). \textsl{Teor\'\i a asint\'otica de estimadores robustos en regresi\'on lineal.} Doctoral thesis, Universidad Nacional de la Plata. Available at \url{http://www.mate.unlp.edu.ar/tesis/tesis\_fasano\_v.pdf}.

\item Glynn, A. and Quinn, K. (2010). An introduction to the augmented inverse propensity weighted estimator. \textsl{Political Analysis}, \textbf{18}, 36-56.

\item {Horvitz, D. G.} and {Thompson, D. J.} (1952). A generalization of sampling without replacement from a finite universe. \textsl{Journal of the American Statistical Association}, \textbf{47}, 663-685.

\item Hristache, M. and Patilea, V. (2017). Conditional moment models with data missing at random. \textsl{Biometrika}, \textbf{104}, 735-742.

\item Huber, P. (1964). Robust estimation of a location parameter. \textsl{Annals of Mathematical Statistics}, \textbf{35}, 73-101.
 
\item Huber, P. and Ronchetti, E.  (2009). \textsl{Robust Statistics}. Wiley, New York, 2nd edition.
 
\item Maronna, R.; Martin,  D.  and Yohai, V.  (2006).  \textsl{Robust Statistics: Theory and Methods} , Wiley, New York.

\item Molina, J.; Sued, M.; Valdora, M. and Yohai, V. (2017). Robust doubly protected estimators for quantiles with missing data. Available at \url{https://arxiv.org/abs/1707.01951}

  \item   M\"uller, U.  (2009). Estimating linear functionals in nonlinear regression with responses missing at random. \textsl{Annals of Statistics}, \textbf{37}, 2245-2277.
  
  \item Pollard, D.  (1984). \textsl{Convergence of Stochastic Processes}. Springer--Verlag, New York.

  \item Robins, J. (1999). Robust estimation in sequentially ignorable missing data and causal inference models.
\textsl{Proceedings of the American Statistical Association Section on Bayesian Statistical Science}, 6-10.

\item Robins, J.; Rotnitzky, A. and Zhao, L. (1994). Estimation of regression coefficients when some regressors are not always observed. \textsl{Journal of the American Statistical Association}, \textbf{89}, 846-866.

\item Scharfstein, D.; Rotnitzky, A. and Robins, J. (1999). Adjusting for non--ignorable drop out in semiparametric non--response models (with discussion). \textsl{Journal of the American Statistical Association}, \textbf{94}, 1096-1146.

\item {Sued, M.} and {Yohai, V.} (2013). Robust location estimation with missing data. \textsl{Canadian Journal of Statistics}, \textbf{41}, 111-132.

 \item Wang, Q. and Qin, Y. (2010). Empirical likelihood confidence bands for distribution functions with missing responses. \textsl{Journal of Statistical Planning and Inference}, \textbf{140}, 2778-2789.
 
 \item Wang, C.; Wang, S.; Zhao, L. and Ou, S. (1997). Weighted semiparametric estimation in regression analysis regression
with missing covariates data. \textsl{Journal of the American Statistical Association}, \textbf{92}, 512-525.

\item  Zhang, Z.; Chen, Z.; Troendle, J. F. and  Zhang, J.(2012) Causal inference on quantiles with an obstetric application.  Biometrics, \textbf{68}, 697-706.
\end{description}

\normalsize

\setcounter{equation}{0}
\renewcommand{\theequation}{A.\arabic{equation}}

\setcounter{section}{0}
\renewcommand{\thesection}{\Alph{section}}
\section{Appendix}{\label{proofs}}

 The following assumptions are needed to derive the asymptotic distribution of the inverse probability weighting and the convolution--based estimators.
 
 \begin{enumerate}
\item[]\textbf{N1.}  The function $\psi$ is twice continuously differentiable with bounded derivatives.
\item[]\textbf{N2.} $A(\psi)=E\left[{\delta}\psi^{\prime}\left(u\right)/{p(\bx,t)}\right]=E\psi^{\prime}\left(u\right)\ne 0$.

\item[]\textbf{N3.} The missingness probability  $p(\bz)=p(\bz,\bgama_0)$, $\bgama_0\in \real^s$, is such that
\begin{itemize}
\item[a)] the family of functions $\itP=\{p(\bz, \bgama): \bgama\in \real^s\}$ has finite entropy.
\item[b)] $p(\bz, \bgama)$ is twice continuously differentiable with respect to $\bgama$. We will denote by $\dot{p}(\bz,\bgama)=(\dot{p}_1(\bz,\bgama), \dots, \dot{p}_s(\bz,\bgama))\trasp$ and $\ddot{p}(\bz,\bgama)=(\ddot{p}_{ij}(\bz,\bgama))$ the gradient and Hessian matrix of $p(\bz, \bgama)$ with respect to $\bgama$.
\item[c)] $\esp\left(\|\dot{p}_j(\bz,\bgama_0)\|/{p(\bz)}\right)<\infty$ for $1\le j\le s$.
\item[d)] For some $\Lambda>0$, $\esp\left(\sup_{\|\bgamach-\bgamach_0\|<\Lambda}\|\ddot{p}_{j\ell}(\bz,\bgama)\| /{p(\bz)}\right)<\infty$ for $1\le j,\ell\le s$.
\end{itemize} 
\item[]\textbf{N4.} $\wbgama$  admits a Bahadur expansion given by $\sqrt{n}\left(\wbgama-\bgama_0\right)=(1/{\sqrt{n}}) \sum_{i=1}^n \etab(\bz_i)+o_\prob(1)$
where  $\esp \etab( \bz)=\bcero$ and $\esp\|\etab( \bz)\|^2<\infty$. We will denote by  $\bSi=\esp\etab( \bz)\etab( \bz)\trasp$ the asymptotic covariance matrix of   $\wbgama$. 
\item[]\textbf{N5.} The missingness probability $p(\bz)$ is {a smooth function of $\bz$, $r-$th continuously differentiable.}
\item[]\textbf{N6.} The bandwidth $b_n$ satisfies that $\rho_n^2=\left\{n b_n^{2r}+(n b_n^{2k})^{-1}\right\}\to 0$
\item[]\textbf{N7.} The kernel $K:\real^{k}\to \real$ is bounded,  has compact support and $\int K(\bu)d\bu>0$. Furthermore, $\int u_j^m K(\bu)d\bu=0$, for $1\le j \le k$, $1\le m\le r-1$,  $\int u_j^{r}\,K(\bu)d\bu>0$, for $1\le j \le k$. 

\item[]\textbf{N8.} The regression function     $\mu(\bx)=m(\bx,\bbe_0)$, $\bbe_0\in \real^{d_1}$, where $d_1$ may not be equal to $d$, is such that
\begin{enumerate}
\item[a)] The function $m(\bx,\bbe)$ is twice continuously differentiable with respect to $\bbe$ and there exists $\eta>0$ such that
$$\esp\sup_{\|\bbech-\bbech_0\|\le \eta} \|\dot{m}(\bx,\bbe)\|^2 <\infty\qquad \esp\sup_{\|\bbech-\bbech_0\|\le \eta} \|\ddot{m}(\bx,\bbe)\|^2 <\infty$$
where $\dot{m}(\bx,\bbe)$ and $\ddot{m}(\bx,\bbe)$ stand for the gradient vector and the Hessian matrix of the function $m(\bx,\bbe)$  with respect to $\bbe$ and for any vector or matrix $\bA$, $\|\bA\|$ denotes its euclidean norm.
\item[b)] The predicted values are computed through $\wmu(\bx)=m(\bx,\wbbe)$, where the estimator $\wbbe$ of $\bbe_0$ admits a Bahadur expansion given by 
$\sqrt{n}\left(\wbbe-\bbe_0\right)=(1/{\sqrt{n}}) \sum_{i=1}^n \delta_i\, \bchi_1(y_i,\bx_i)+o_\prob(1)$,
with  $\esp \bchi_1(y,\bx)=\bcero$ and $\esp\|\bchi_1(y,\bx)\|^2<\infty$. 

\end{enumerate}
 
\end{enumerate}

The proof of Theorem {\ref{ws}}.1 is omitted since it follows using  analogous arguments to those considered in Theorems 4.1 to 4.3 in  Bianco \textsl{et al.} (2011).  

\noi \textsc{Proof of Theorem {\ref{conv}}.1.} Recall that $\wy_{ij} =  y_i-\wmu(\bx_i)+  \wmu(\bx_j)  $ for  $i, j \in \{\delta_\ell=1\}$, $\tau_j=\left\{\sum_{\ell=1}^n  { \delta_{\ell}}/{\wpe(\bz_\ell)}\right\}^{-1}  {\delta_{j}}/{\wpe(\bz_j)}$ and $\kappa_i=\left\{\sum_{\ell=1}^n \delta_{\ell}\right\}^{-1}     { \delta_{i}}$. For simplicity, denote as $\wtheta$,  the solution of $\wlam_{\conv}(\wpe, \wvarsigma, a)=0$.  Then, we have that
\begin{eqnarray*}
0&=&  \sum_{i=1}^n\sum_{j=1}^n \tau_j \kappa_i \psi \left(\frac{\wy_{ij}-\wtheta}{\wvarsigma}\right)=  \sum_{i=1}^n\sum_{j=1}^n \tau_j \kappa_i \psi \left(\frac{\wy_{ij}-\theta}{\wvarsigma}\right) + \frac{\theta-\wtheta}{\wvarsigma}\,A_n
\end{eqnarray*}
where 
$$A_n=\sum_{i=1}^n\sum_{j=1}^n \tau_j \kappa_i \psi^{\prime}  \left(\frac{\wy_{ij}-\theta}{\wvarsigma}\right)  + \frac{(\wtheta-\theta) }{2\, \wvarsigma }\,  \sum_{i=1}^n\sum_{j=1}^n \tau_j \kappa_i \psi^{\prime\prime} \left(\frac{\wy_{ij}-\wnu}{\wvarsigma}\right) =A_{1,n}+A_{2,n}\,,$$
with $\wnu$ is an intermediate point between $\theta$ and $\wtheta$.
 Hence,
$$A_n\;\frac{\wtheta-\theta}{\wvarsigma}=\sum_{i=1}^n\sum_{j=1}^n \tau_j \kappa_i \psi \left(\frac{\wy_{ij}-\theta}{\wvarsigma}\right)\,.$$
Using that $\sum_i \kappa_i=\sum_j \tau_j=1$, we get that
$|A_{2,n}|\le  {|\wtheta-\theta| }\, \|\psi^{\prime\prime}\|_{\infty}/({2\, \wvarsigma }\,)$
which together with the consistency of $\wtheta$ and $\wvarsigma$ entail that $A_{2,n}\convprob 0$. 

Note that $A_{1,n}$ can be written as
$A_{1,n}=M(\wQ_{y,\conv}, \wvarsigma\,)$, where 
$M(Q,\varsigma)=\int \psi^{\prime}  \left(({y-\theta})/{\varsigma}\right) dQ(y)$.
The fact that $t\psi^{\prime}(t)$ is bounded allows to show easily that $M(\wQ_{y,\conv}, \wvarsigma)- M(\wQ_{y,\conv}, \varsigma_0) \convprob 0$, since $\wvarsigma\convprob \varsigma_0$. On the other hand, using that $\psi^{\prime}$ is bounded we obtain that the functional $M^{\star}(Q)=M(Q,\varsigma_0)$ is continuous with respect to the Prohorov distance. Therefore, using that $\Pi(\wQ_{y, \conv} ,Q_y)\convpp 0$, we obtain that $M(\wQ_{y, \conv},\varsigma_0) \convpp M(Q_y,\varsigma_0)$, which together with the fact that $M(\wQ_{y,\conv}, \wvarsigma)- M(\wQ_{y,\conv}, \varsigma_0) \convprob 0$ and $A_{2,n}\convprob 0$ leads us to 
$$ 
A_{n} \convprob \esp \psi^{\prime}\left( \frac{y-\theta}{\varsigma_0}\right)=\esp \psi^{\prime}\left(u\right)=A\ne 0 \,,$$ 
so
$ 
\sqrt{n}(\wtheta-\theta)= \varsigma_0 A^{-1} B_n +o_\prob(1)$
where
\begin{equation}
B_n=\sqrt{n} \sum_{i=1}^n\sum_{j=1}^n \tau_j \kappa_i \psi \left(\frac{\wy_{ij}-\theta}{\wvarsigma}\right)\,.
\label{Bn}
\end{equation}

Using that  $\mu(\bx)=m(\bx,\bbe_0)$ and $\wmu(\bx)=m(\bx,\wbbe)$ and denoting $y_{ij}=y_i-\mu(\bx_i)+  \mu(\bx_j)=\epsilon_i+  \mu(\bx_j)$ and $\wDelta(\bx)=\wmu(\bx)-\mu(\bx)$, we get that
$$\wy_{ij}=y_i-\wmu(\bx_i)+  \wmu(\bx_j) =\epsilon_i + \mu(\bx_i)-\wmu(\bx_i)+  \wmu(\bx_j) =y_{ij}+ \wDelta(\bx_j)-\wDelta(\bx_i)\,,$$
which implies that
\begin{eqnarray*}
\psi \left(\frac{\wy_{ij}-\theta}{\wvarsigma}\right)&=& \psi \left(\frac{y_{ij}-\theta}{\wvarsigma}\right)+ \psi^{\prime} \left(\frac{y_{ij}-\theta}{\wvarsigma}\right) \frac{\wDelta(\bx_j)-\wDelta(\bx_i)}{\wvarsigma}+  \frac 12 \psi^{\prime\prime} \left(\frac{y_{ij}-\wtnu_{ij}}{\wvarsigma}\right) \frac{\left(\wDelta(\bx_j)-\wDelta(\bx_i)\right)^2}{\wvarsigma^2}
\end{eqnarray*}
with $\wtnu_{ij}$ an intermediate point between $0$ and  $\wDelta(\bx_j)-\wDelta(\bx_i)$. Hence, from (\ref{Bn}) we have that $B_n=\sum_{i=1}^3 B_{n,i}$ where
\begin{eqnarray*}
B_{n,1}&=& \sqrt{n} \sum_{i=1}^n\sum_{j=1}^n \tau_j \kappa_i  \psi \left(\frac{y_{ij}-\theta}{\wvarsigma}\right)\\
B_{n,2}&=& \frac 1{\wvarsigma} \,\sqrt{n} \sum_{i=1}^n\sum_{j=1}^n \tau_j \kappa_i  \psi^{\prime} \left(\frac{y_{ij}-\theta}{\wvarsigma}\right) \left(\wDelta(\bx_j)-\wDelta(\bx_i)\right)\\
B_{n,3}&=&  \frac 1{2\, \wvarsigma^2} \sqrt{n} \sum_{i=1}^n\sum_{j=1}^n \tau_j \kappa_i \psi^{\prime\prime} \left(\frac{y_{ij}-\wtnu_{ij}}{\wvarsigma}\right)  \left(\wDelta(\bx_j)-\wDelta(\bx_i)\right)^2 \,.
\end{eqnarray*}
Using that $\left(\wDelta(\bx_j)-\wDelta(\bx_i)\right)^2\le 2\left( \wDelta^2(\bx_j)+\wDelta^2(\bx_i)\right)$ and $\wDelta(\bx)=\wmu(\bx)-\mu(\bx)=\dot{m}(\bx,\wtbbe)\trasp (\wbbe -\bbe _0)$ for some $\wtbbe$ between $\wbbe$ and $\bbe _0$, we obtain that
\begin{eqnarray*}
|B_{n,3}| & \le &  \frac { \|\psi^{\prime\prime}\|_{\infty}}{  \wvarsigma^2} \sqrt{n}\left\{ \sum_{j=1}^n \tau_j \wDelta^2(\bx_j) +\sum_{i=1}^n \kappa_i  \wDelta^2(\bx_i)\right\} \\
& \le  &\frac { \|\psi^{\prime\prime}\|_{\infty}}{  \wvarsigma^2} \sqrt{n}(\wbbe -\bbe _0)\trasp \left\{ \sum_{j=1}^n \tau_j    \dot{m}(\bx_j,\wtbbe_j) \dot{m}(\bx_j,\wtbbe_j)\trasp  +\sum_{i=1}^n \kappa_i  \dot{m}(\bx_i,\wtbbe_i) \dot{m}(\bx_i,\wtbbe_i)\trasp\right\} (\wbbe -\bbe _0) \,.
\end{eqnarray*}
  From \textbf{N8}a), we have that $\esp\sup_{\|\bbech-\bbech_0\|\le \eta} \|\dot{m}(\bx_1,\bbe)\|^2 <\infty$, so  the consistency of $\wbbe$ together with \textbf{A1} and  $\sup_{\bz\in \itS_\bz}|\wpe(\bz)-p(\bz)|\convprob 0$ (when $\wpe(z)=p(\bz,\wbgama)$) imply that $\sum_{i=1}^n \kappa_i  \dot{m}(\bx_i,\wtbbe_i) \dot{m}(\bx_i,\wtbbe_i)\trasp=O_\prob(1)$. On the other hand, \textbf{N8}b) implies that $\sqrt{n}(\wbbe -\bbe _0)=O_\prob(1)$. Thus, $B_{n,3}\convprob 0$ and $B_n= B_{n,1}+B_{n,2}+o_\prob(1)$. We now expand $B_{n,2}$ as $B_{n,2}=B_{n,2,1}+B_{n,2,2}$ with
  \begin{eqnarray*}
 B_{n,2,1} &=&  \frac 1{\wvarsigma} \,\sqrt{n} (\wbbe -\bbe _0)\trasp \sum_{i=1}^n\sum_{j=1}^n \tau_j \kappa_i  \psi^{\prime} \left(\frac{y_{ij}-\theta}{\wvarsigma}\right) \left(\dot{m}(\bx_j,\bbe_0)- \dot{m}(\bx_i,\bbe_0) \right) \\
 B_{n,2,2}&=&   \frac 1{2\wvarsigma} \,\sqrt{n}  (\wbbe -\bbe _0)\trasp\, \sum_{i=1}^n\sum_{j=1}^n \tau_j \kappa_i  \psi^{\prime} \left(\frac{y_{ij}-\theta}{\wvarsigma}\right)\left(\ddot{m}(\bx_j,\wtbbe_j)- \ddot{m}(\bx_i,\wtbbe_i) \right)\trasp (\wbbe -\bbe _0) 
\end{eqnarray*}
Using that $\sqrt{n}  (\wbbe -\bbe _0)=O_\prob(1)$, $\psi^{\prime}$ is bounded and $\esp\sup_{\|\bbech-\bbech_0\|\le \eta} \|\ddot{m}(\bx_1,\bbe)\|^2 <\infty$, we get easily that $B_{n,2,2}\convprob 0$. Hence, $B_n$ has the same asymptotic behaviour as $B_{n,1}+B_{n,2,1}$.

The fact that  $\sqrt{n}  (\wbbe -\bbe _0)=O_\prob(1)$, $\wvarsigma\convprob \varsigma_0$, together with \textbf{N8}a) entails that
$$B_{n,,2,1}-\frac 1{\wvarsigma} \,\sqrt{n} (\wbbe -\bbe _0)\trasp \sum_{i=1}^n\sum_{j=1}^n \tau_j \kappa_i  \psi^{\prime} \left(\frac{y_{ij}-\theta}{\varsigma_0}\right) \left(\dot{m}(\bx_j,\bbe_0)- \dot{m}(\bx_i,\bbe_0) \right) \convprob 0\,.$$
Denote as
\begin{eqnarray*}
\bC_{n,1}&=& \sum_{i=1}^n\sum_{j=1}^n \tau_j \kappa_i  \psi^{\prime} \left(\frac{y_{ij}-\theta}{\varsigma_0}\right)  \dot{m}(\bx_j,\bbe_0) \\
\bC_{n,2}&=& \sum_{i=1}^n\sum_{j=1}^n \tau_j \kappa_i  \psi^{\prime} \left(\frac{y_{ij}-\theta}{\varsigma_0}\right)  \dot{m}(\bx_i,\bbe_0) \,.
\end{eqnarray*}
Then,
$$B_n= B_{n,1}+ \frac 1{\wvarsigma} \,\sqrt{n} (\wbbe -\bbe _0)\trasp\left(\bC_{n,1}-\bC_{n,2}\right)+o_{\prob}(1)\,.$$
Using \textbf{A1} and the fact that $\sup_{\bz\in \itS_\bz}|\wpe(\bz)-p(\bz)|\convprob 0$, when $\wpe(z)=p(\bz,\wbgama)$, it is easy to show that
$$\bC_{n,2}\convprob \bC_2=\frac{1}{\esp \delta_1}\esp p(\bz_1)  \psi^{\prime} \left(\frac{y_{12}-\theta}{\varsigma_0}\right)  \dot{m}(\bx_1,\bbe_0)=\frac{1}{\esp \delta_1}\esp p(\bz_1)  \psi^{\prime} \left(\frac{\epsilon_1+m(\bx_2, \bbe_0)-\theta}{\varsigma_0}\right)  \dot{m}(\bx_1,\bbe_0)$$
Similarly,
$$\bC_{n,1}\convprob \bC_1=\frac{1}{\esp \delta_1}\esp p(\bz_1)  \psi^{\prime} \left(\frac{y_{12}-\theta}{\varsigma_0}\right)  \dot{m}(\bx_2,\bbe_0)=\frac{1}{\esp \delta_1}\esp p(\bz_1)  \psi^{\prime} \left(\frac{\epsilon_1+m(\bx_2, \bbe_0)-\theta}{\varsigma_0}\right)  \dot{m}(\bx_2,\bbe_0)$$
leading to
$$\bC_n=\bC_{n,1}-\bC_{n,2}\convprob  \bC=\bC_1-\bC_2=\frac{1}{\esp \delta_1}\esp p(\bz_1)  \psi^{\prime} \left(\frac{y_{12}-\theta}{\varsigma_0}\right)  \left\{\dot{m}(\bx_2,\bbe_0) -  \dot{m}(\bx_1,\bbe_0)\right\}$$
 Therefore, using that  $\sqrt{n}  (\wbbe -\bbe _0)=O_\prob(1)$ and $\wvarsigma\convprob \varsigma_0$, we obtain that
 $$B_n= \sqrt{n} \sum_{i=1}^n\sum_{j=1}^n \tau_j \kappa_i  \psi \left(\frac{y_{ij}-\theta}{\wvarsigma}\right)+ \frac 1{\varsigma_0} \,\sqrt{n} (\wbbe -\bbe _0)\trasp \bC + o_{\prob}(1)\,.$$
 
Recall that, since the errors and the covariates are independent, we have that $y_{12}\sim y_1$ which together with the fact that $\esp\psi ((y_1-\theta)/\varsigma)=0$ for all $\varsigma>0$  allows to show, using standard empirical process arguments, that
 $$\sqrt{n} \sum_{i=1}^n\sum_{j=1}^n \tau_j \kappa_i  \psi \left(\frac{y_{ij}-\theta}{\wvarsigma}\right)- \sqrt{n} \sum_{i=1}^n\sum_{j=1}^n \tau_j \kappa_i  \psi \left(\frac{y_{ij}-\theta}{\varsigma_0}\right)\convprob 0\,.$$
 Hence
 $$B_n= \sqrt{n} \sum_{i=1}^n\sum_{j=1}^n \tau_j \kappa_i  \psi \left(\frac{y_{ij}-\theta}{\varsigma_0}\right)+ \frac 1{\varsigma_0} \,\sqrt{n} (\wbbe -\bbe _0)\trasp \bC + o_{\prob}(1)\,.$$
 We will now use that $\sqrt{n}\left(\wbbe-\bbe_0\right)=(1/{\sqrt{n}}) \sum_{i=1}^n \delta_i\, \bchi_1(y_i,\bx_i)+o_\prob(1)$ obtaining that
$$
B_n= \sqrt{n} \sum_{i=1}^n\sum_{j=1}^n \tau_j \kappa_i  \psi \left(\frac{y_{ij}-\theta}{\varsigma_0}\right)+ \frac 1{\varsigma_0} \,\frac 1{\sqrt{n}}  \sum_{i=1}^n \delta_i\, \bchi_1(y_i,\bx_i) \trasp \bC + o_{\prob}(1)\,.
$$
Recall that $\tau_j=\left(\sum_{\ell=1}^n  { \delta_{\ell}}/{\wpe(\bz_\ell)}\right)^{-1}  {\delta_{j}}/{\wpe(\bz_j)}$ and $\kappa_i=\left\{\sum_{\ell=1}^n \delta_{\ell}\right\}^{-1}     { \delta_{i}}$, so using  again \textbf{A1} and the fact that $\sup_{\bz\in \itS_\bz}|\wpe(\bz)-p(\bz)|\convprob 0$, when $\wpe(z)=p(\bz,\wbgama)$, we get that $\sum_{\ell=1}^n \delta_{\ell}/n\convprob \esp \delta_1$ and $\sum_{\ell=1}^n  { \delta_{\ell}}/{\wpe(\bz_\ell)}\convprob 1$, leading to
\begin{equation}
B_n= \frac{1}{\esp \delta_1}\left\{\sqrt{n} \frac{1}{n^2}\sum_{i=1}^n\sum_{j=1}^n  \frac{\delta_{j}}{\wpe(\bz_j)}\; \delta_i  \psi \left(\frac{y_{ij}-\theta}{\varsigma_0}\right)+ \frac 1{\varsigma_0} \,\frac 1{\sqrt{n}}  \sum_{i=1}^n \delta_i\, \bchi_1(y_i,\bx_i) \trasp \bC^{\star} \right\}+ o_{\prob}(1)\,,
\label{desarrolloBn}
\end{equation}
with $\bC^{\star}=\esp \delta_1\;\bC$.

\noi a) Let us begin by considering that $\wpe(\bz)=p(\bz)$. In this case,
$$B_n= \frac{1}{\esp \delta_1}\left\{\sqrt{n} \frac{1}{n^2}\sum_{i=1}^n\sum_{j=1}^n  \frac{\delta_{j}}{p(\bz_j)}\; \delta_i  \psi \left(\frac{y_{ij}-\theta}{\varsigma_0}\right)+ \frac 1{\varsigma_0} \,\frac 1{\sqrt{n}}  \sum_{i=1}^n \delta_i\, \bchi_1(y_i,\bx_i) \trasp \bC^{\star} \right\}+ o_{\prob}(1) $$
 and the result follows using standard $U-$statistics arguments.  
 Effectively, 
 $$\psi \left(\frac{y_{ij}-\theta}{\varsigma_0}\right)= \psi \left(\frac{\epsilon_i+\mu(\bx_j)-\theta}{\varsigma_0}\right)$$
 therefore, using that $\epsilon_i+\mu(\bx_j)\sim \epsilon_i+\mu(\bx_i)$, we get that for $\ell\ne i$ and $\ell\ne j$
 $$\esp \left[ \frac{\delta_{j}}{p(\bz_j)}\; \delta_i \psi \left(\frac{y_{ij}-\theta}{\varsigma_0}\right)|(\epsilon_\ell, \bx_\ell, \delta_\ell)\right]=\esp \left[ \frac{\delta_{j}}{p(\bz_j)}\; \delta_i \psi \left(\frac{\epsilon_i+\mu(\bx_j)-\theta}{\varsigma_0}\right)|(\epsilon_\ell, \bx_\ell, \delta_\ell)\right]=0\,.$$
 Hence, straightforward calculations allow to show that 
 $$\esp \left[\sum_{i=1}^n\sum_{j=1}^n \frac{\delta_{j}}{p(\bz_j)}\; \delta_i \psi \left(\frac{y_{ij}-\theta}{\varsigma_0}\right)|(\epsilon_\ell, \bx_\ell, \delta_\ell)\right]=\frac{\delta_{\ell}}{p(\bz_\ell)}  \psi \left(\frac{y_{\ell}-\theta}{\varsigma_0}\right) + (n-1) A(\bx_\ell, \delta_\ell) + (n-1) B(\epsilon_\ell,\delta_\ell) $$ 
where $A(\bx_0, \delta_0)$ and $B(\epsilon_0, \delta_0)$ are defined in (\ref{Ax0}) and (\ref{Beps0}), respectively.
Hence, if we denote as
\begin{eqnarray*}
V_n&=&\frac{1}{\esp \delta_1}\left\{\sqrt{n} \frac{1}{n^2}\sum_{\ell=1}^n  \left[\frac{\delta_{\ell}}{p(\bz_\ell)}  \psi \left(\frac{y_{\ell}-\theta}{\varsigma_0}\right) + (n-1) A(\bx_\ell, \delta_\ell) + (n-1) B(\epsilon_\ell,\delta_\ell)\right]  \right. \\
&& \left.+ \frac 1{\varsigma_0} \,\frac 1{\sqrt{n}}  \sum_{i=1}^n \delta_i\, \bchi_1(y_i,\bx_i) \trasp \bC^{\star} \right\}
\end{eqnarray*}
we have that $B_n-V_n=o_\prob(1)$. Note that
$$\left|\sqrt{n} \frac{1}{n^2}\sum_{\ell=1}^n   \frac{\delta_{\ell}}{p(\bz_\ell)}  \psi \left(\frac{y_{\ell}-\theta}{\varsigma_0}\right)\right|\le \|\psi\|_{\infty} \inf_{\bz\in \itS_\bz}p(\bz) \frac{1}{\sqrt{n}}$$
so, 
$$B_n=\frac{1}{\esp \delta_1}\left\{\frac 1{\sqrt{n}}\sum_{\ell=1}^n  \left[   A(\bx_\ell, \delta_\ell) +   B(\epsilon_\ell,\delta_\ell)  + \frac 1{\varsigma_0} \, \delta_\ell\, \bchi_1(y_\ell,\bx_\ell) \trasp \bC^{\star}\right]  \right\}+o_\prob(1)$$
and the result follows from the Central Limit Theorem.

 \noi b) We will now consider the situation in which $\wpe(\bz)=p(\bz,\wbgama)$. From (\ref{desarrolloBn}) we have that
 $$B_n=\frac{1}{\esp \delta_1}\left\{ D_n(\wpe)+ \frac 1{\varsigma_0} \,\frac 1{\sqrt{n}}  \sum_{i=1}^n \delta_i\, \bchi_1(y_i,\bx_i) \trasp \bC^{\star} \right\}+ o_{\prob}(1) $$
 where
 $$D_n(q)= \sqrt{n} \frac{1}{n^2}\sum_{i=1}^n\sum_{j=1}^n  \frac{\delta_{j}}{q(\bz_j)}\; \delta_i  \psi \left(\frac{y_{ij}-\theta}{\varsigma_0}\right)\,. $$
 Note that $D_n(\wpe)=D_n(p)+W_n$ with 
 $$W_n=\sqrt{n} \frac{1}{n^2}\sum_{i=1}^n\sum_{j=1}^n  \frac{\delta_{j}}{\wpe(\bz_j)\,p(\bz_j)}\;\left[p(\bz_j)-\wpe(\bz_j)\right] \delta_i  \psi \left(\frac{y_{ij}-\theta}{\varsigma_0}\right)$$
 hence, arguing as in a) we get that
\begin{equation}
B_n=\frac{1}{\esp \delta_1}\left\{\frac 1{\sqrt{n}}\sum_{\ell=1}^n  \left[   A(\bx_\ell, \delta_\ell) +   B(\epsilon_\ell,\delta_\ell)  + \frac 1{\varsigma_0} \, \delta_\ell\, \bchi_1(y_\ell,\bx_\ell) \trasp \bC^{\star}\right]  \right\}+ \frac{1}{\esp \delta_1} W_n +o_\prob(1)\,.
\label{desarrollo2}
\end{equation}
 We will expand $W_n$. For that purpose, define
 $H(\bz,\bgama)= {p(\bz_j, \bgama_0)}/{p(\bz_j, \bgama)}$.
 Then, the gradient $\dot{H}(\bz,\bgama)$ and Hessian $\ddot{H}(\bz,\bgama)$ of $H$ with respect to $\bgama$ are given by
 $$\dot{H}(\bz,\bgama)= -  \frac{p(\bz, \bgama_0)}{p^2(\bz,\bgama)}\, \dot{p}(\bz,\bgama) \qquad \ddot{H}(\bz,\bgama) =  -  \frac{p(\bz, \bgama_0)}{p^2(\bz,\bgama)}\, \ddot{p}(\bz,\bgama)+ 2     \frac{p(\bz, \bgama_0)}{p^3(\bz,\bgama)}\, \dot{p}(\bz,\bgama)\dot{p}(\bz,\bgama)\trasp\,.$$
 In particular, we have that
  $\dot{H}(\bz,\bgama_0)= -  \, \dot{p}(\bz,\bgama_0)/{p(\bz, \bgama_0)}=-  \, \dot{p}(\bz,\bgama_0)/{p(\bz)}$.
  Therefore, using a Taylor's expansion we get that
\begin{eqnarray*}
W_n
&=&\sqrt{n} (\wbgama-\bgama_0)\trasp  \frac{1}{n^2}\sum_{i=1}^n\sum_{j=1}^n  \frac{\delta_{j}}{p(\bz_j)}\;\dot{H}(\bz_j, \bgama_0) \delta_i  \psi \left(\frac{y_{ij}-\theta}{\varsigma_0}\right)\\
&&+\frac 12 \sqrt{n} (\wbgama-\bgama_0)\trasp \frac{1}{n^2}\sum_{i=1}^n\sum_{j=1}^n  \frac{\delta_{j}}{p(\bz_j)}\; \ddot{H}(\bz_j, \widetilde{\bgama}) \delta_i  \psi \left(\frac{y_{ij}-\theta}{\varsigma_0}\right) (\wbgama-\bgama_0)\,,
\end{eqnarray*}
where $\widetilde{\bgama}$ is an intermediate point between $\bgama_0$ and $\wbgama$. Hence,
 $W_n=
\,-\,\sqrt{n} (\wbgama-\bgama_0)\trasp\, \bW_{n,1}+W_{n,2}$,
where
\begin{eqnarray*}
\bW_{n,1}
&=&  \frac{1}{n^2}\sum_{i=1}^n\sum_{j=1}^n  \frac{\delta_{j}}{p^2(\bz_j)}\;  \dot{p}(\bz_j,\bgama_0)  \delta_i  \psi \left(\frac{y_{ij}-\theta}{\varsigma_0}\right)\\
W_{n,2}&=&\frac 12 \sqrt{n} (\wbgama-\bgama_0)\trasp \frac{1}{n^2}\sum_{i=1}^n\sum_{j=1}^n  \frac{\delta_{j}}{p(\bz_j)}\; \ddot{H}(\bz_j, \widetilde{\bgama}) \delta_i  \psi \left(\frac{y_{ij}-\theta}{\varsigma_0}\right) (\wbgama-\bgama_0) \,.
\end{eqnarray*}
The fact that $W_{n,2}$ may be bounded as
 $$2 \, |W_{n,2}|\le \|\psi\|_{\infty}\, \sqrt{n} \|\wbgama-\bgama_0\|^2 \frac{1}{n}\sum_{j=1}^n  \frac{1}{p(\bz_j)} \|\ddot{H}(\bz_j, \widetilde{\bgama}) \|\,,
$$ 
together with \textbf{N3}d) and \textbf{N4}, imply that $W_{n,2}\convprob 0$.

To obtain an expansion for $ \sqrt{n} (\wbgama-\bgama_0)\trasp \bW_{n,1}$ we define
$$r_{i,j}(\bz_0)= \esp\left[\psi \left(\frac{y_{ij}-\theta}{\varsigma_0}\right)|\bz_j=\bz_0\right]=\esp\left[\psi \left(u_i+\frac{\mu(\bx_j)-\mu(\bx_i)}{\varsigma_0}\right)|\bz_j=\bz_0\right]\,.$$
Then, $r_{i,i}(\bz_0)=r(\bz_0)$, while for $i\ne j$, $r_{i,j}(\bz_0)=r_{1,2}(\bz_0)$. Therefore, 
\begin{eqnarray*}
\bW_{n,1} 
&=& \frac{1}{n^2}\sum_{i=1}^n\sum_{j=1}^n  \frac{\delta_{j}}{p^2(\bz_j)}\;  \dot{p}(\bz_j,\bgama_0)  \delta_i  \left[\psi \left(\frac{y_{ij}-\theta}{\varsigma_0}\right)- r_{i,j}(\bz_j)\right]\\
&& +  \frac{1}{n}\sum_{j=1}^n  \frac{\delta_{j}}{p^2(\bz_j)}\;  \dot{p}(\bz_j,\bgama_0)     r_{1,2}(\bz_j) \left( \frac 1n\sum_{i=1}^n \delta_i \right)  +  \frac{1}{n^2}\sum_{i=1}^n  \frac{\delta_{i}}{p^2(\bz_i)}\;  \dot{p}(\bz_i,\bgama_0)    \left[  r(\bz_i) -r_{1,2}(\bz_i)\right]\,.
\end{eqnarray*}
It is easy to see that
$$\frac{1}{n^2}\sum_{i=1}^n\sum_{j=1}^n  \frac{\delta_{j}}{p^2(\bz_j)}\;  \dot{p}(\bz_j,\bgama_0)  \delta_i  \left[\psi \left(\frac{y_{ij}-\theta}{\varsigma_0}\right)- r_{i,j}(\bz_j)\right]\convprob 0\,,$$
since it is a centered $U-$statistic. On the other hand, we also have that
$$\frac{1}{n^2}\sum_{i=1}^n  \frac{\delta_{i}}{p^2(\bz_i)}\;  \dot{p}(\bz_i,\bgama_0)    \left[  r(\bz_i) -r_{1,2}(\bz_i)\right]\convprob 0\,,$$
while 
$$ \frac{1}{n}\sum_{j=1}^n  \frac{\delta_{j}}{p^2(\bz_j)}\;  \dot{p}(\bz_j,\bgama_0)     r_{1,2}(\bz_j) \left( \frac 1n\sum_{i=1}^n \delta_i \right)  \convprob \esp (\delta_1)\,  \esp (1/{p(\bz_1)})\;  \dot{p}(\bz_1,\bgama_0)     r_{1,2}(\bz_1)= \esp(\delta_1) \bd_1$$
Therefore, using \textbf{N4}, we get that
$$W_n= \,-\,\sqrt{n} (\wbgama-\bgama_0)\trasp\,\esp(\delta_1) \bd_1 + o_\prob(1)= \,-\, \frac{1}{\sqrt{n}}\sum_{i=1}^n \etab(\bz_i)\trasp\,\esp(\delta_1) \bd_1 + o_\prob(1)$$
which together with (\ref{desarrollo2}) leads to
$$
B_n=\frac{1}{\esp \delta_1}\left\{\frac 1{\sqrt{n}}\sum_{\ell=1}^n  \left[   A(\bx_\ell, \delta_\ell) +   B(\epsilon_\ell,\delta_\ell)  + \frac 1{\varsigma_0} \, \delta_\ell\, \bchi_1(y_\ell,\bx_\ell) \trasp \bC^{\star}\,-\,\esp(\delta_1) \,   \etab(\bz_\ell)\trasp\, \bd_1\right]  \right\}+  o_\prob(1)\,.$$
and the result follows from the Central Limit Theorem. \square
 
 \vskip0.1in
 \noi \textsc{Proof of Theorem \ref{sec:DR}.1.} Denote as $\pi_1=\pi(\bz_1)$ and note that $Q_y=\nu_1+\nu_2$, where for any borelian set $A$,
\begin{eqnarray*}
\nu_1(A)=\esp \left(\frac{\delta_{1}}{\pi_{1}} \Delta_{y_1}(A)\right) & \qquad\mbox{and} \qquad & 
\nu_2 (A) =   \esp \left\{\left(1-\frac{\delta_{1}}{\pi_{1}}\right)  \prob(y\in A|\bz) \right\}\,.
\end{eqnarray*}
Write $\wQ_{y,\doubrob}=\wgamma_1\wQ_{y,1} + \wnu_{2}$, where
\begin{eqnarray*}
\wgamma_1&=& \frac 1n \sum_{i=1} ^n \frac{\delta_{i}}{\wpi_{i}}\qquad  
\wQ_{y,1}= \left\{\sum_{i=1} ^n \frac{\delta_{i}}{\wpi_{i}}\right\}^{-1} \sum_{i=1}^n \frac{\delta_{i}}{\wpi_{i}} \Delta_{ y_i } \quad
\wnu_{2}\left((-\infty,y]\right)= \frac 1n    \sum_{i=1}^n  \left(1-\frac{\delta_{i}}{\wpi(\bz_i)}\right) G_n(y|\bz_i)\,.
\end{eqnarray*}
Similar arguments to those considered in Theorem 3.1 in Bianco \textsl{et al.} (2018) allow to show that $\Pi_{\infty}(\wQ_{y,1},\Upsilon_1)\convpp 0$ and $\Pi(\wQ_{y,1},\Upsilon_1)\convpp 0$, where the probability measure $\Upsilon_1$ is given by $\Upsilon_1= \gamma_1^{-1}\nu_1$ and $\Pi$ and $\Pi_{\infty}$ stand for the Prohorov and Kolmogorov distance respectively. On the other hand, from the Strong Law of Large Numbers we get that $\wgamma_1\convpp \gamma_1=\esp(\delta_1/\pi(\bx_1))=\esp(p(\bx_1)/\pi(\bx_1))>0$. Hence,   $\Pi_{\infty}(\wgamma_1\wQ_{y,1}, \nu_1)\convpp 0$, i.e., 
\begin{equation}
\|\wgamma_1\wQ_{y,1}\left((-\infty,\cdot]\right) -\nu_1\left((-\infty,\cdot]\right)\|_{\infty}\,.
\label{parte1}
\end{equation} 
Furthermore, standard arguments as those considered in the proof of Proposition 3.2.1 in Boente \textsl{et al.} (2009) allow to show that, for any compact set $\itC$, we have  
$\sup_{y\in \real} \sup_{\bz \in \itC} |G_n(y|\bz)-G(y|\bz)|\convpp 0$, which easily entails that 
\begin{equation}
\sup_{y\in \real} |\wnu_{2}\left((-\infty,y]\right)-\nu_2\left((-\infty,y]\right)|\convpp 0\,.
\label{parte2}
\end{equation} 
Combining (\ref{parte1}) and (\ref{parte2}), we obtain  that $\|\wF_{y,\aipw}-F_y\|_\infty\convpp 0$, since for any $\epsilon>0$ there exists a compact set $\itC$ such that $\prob(\bz_1\in \itC)>1-\epsilon$.

Furthermore, similar arguments to those considered in Lemma 1 in Bianco and Boente (2004) allow to see that for any borelian set $A$, $ |\wnu_{2}\left(A\right)-\nu_2\left(A\right)|\convpp 0$, which together with the fact that $\Pi(\wQ_{y,1},\Upsilon_1)\convpp 0$ and $\wgamma_1\convpp \gamma_1$ entails that $|\wQ_{y,\aipw}(A)-Q_y(A)|\convpp 0$. The conclusion now follows from Lemma 7.1 in Bianco \textsl{et al.} (2018).  \square

\end{document}